\documentclass[12pt,epsfig,onecolumn,draftclsnofoot]{IEEEtran}

\usepackage{graphicx}
\usepackage{amssymb}
\usepackage{cite}
\usepackage{amsmath}
\usepackage{multirow}
\usepackage{subfig}
\usepackage{stmaryrd}
\usepackage{comment}
\usepackage{epstopdf}
\usepackage{color}
\usepackage{algorithm,algpseudocode}
\graphicspath{{figures/}}

\newcommand{\bm}{\boldsymbol}

\newtheorem{theorem}{Theorem}
\newtheorem{remark}{Remark}
\newtheorem{assumption}{Assumption}

\begin{document}

\title{Line Spectral Estimation with Unlimited Sensing}

\author{Hongwei Wang, Jun Fang, Hongbin Li,~\IEEEmembership{Fellow,~IEEE},
and Geert Leus,~\IEEEmembership{Fellow,~IEEE}
\thanks{Hongwei Wang, and Jun Fang are with the National Key Laboratory
of Wireless Communications, University of
Electronic Science and Technology of China, Chengdu 611731, China,
Email: hongwei\_wang@uestc.edu.cn, JunFang@uestc.edu.cn.}
\thanks{Hongbin Li is
with the Department of Electrical and Computer Engineering,
Stevens Institute of Technology, Hoboken, NJ 07030, USA, E-mail:
Hongbin.Li@stevens.edu.}
\thanks{Geert Leus is with the Faculty of Electrical Engineering,
Mathematics and Computer Science, Delft University of Technology,
2826 CD Delft, The Netherlands, Email: g.j.t.leus@tudelft.nl.}
}

\maketitle

%The recently introduced unlimited sensing framework (USF) is a
%promising approach to deal with the information loss caused by the
%clipping and saturation in the conventional analog-to-digital
%converters (ADC) due to its limited dynamic range (DR).

\begin{abstract}
In the paper, we consider the line spectral estimation problem in
an unlimited sensing framework (USF), where a modulo
analog-to-digital converter (ADC) is employed to fold the input
signal back into a bounded interval before quantization. Such
an operation is mathematically equivalent to taking the modulo of
the input signal with respect to the interval. To overcome the
noise sensitivity of higher-order difference-based methods, we
explore the properties of the first-order difference of modulo
samples, and develop two line spectral estimation algorithms based on first-order difference, which are robust against noise. Specifically, we show that, with a high probability, the
first-order difference of the original samples is equivalent to
that of the modulo samples. By utilizing
this property, line spectral estimation is solved via a robust
sparse signal recovery approach. The second algorithms is built on our finding that, with a sufficiently high sampling rate, the first-order difference of the original samples can be decomposed as a sum of the first-order difference of the modulo samples and a sequence whose elements are confined to be three possible values. This decomposition enables us to formulate the line spectral estimation problem as a mixed integer linear program that can be efficiently solved. Simulation results show that both proposed methods are robust against noise and achieve a significant performance improvement over the higher-order difference-based method.
\end{abstract}


%ultra wideband channel estimation in wireless
%communications~\cite{BajwaHaupt10},

\section{Introduction}
Line spectral estimation (LSE) is a fundamental problem
in statistical signal processing, which is aimed to detect and
identify the components of a mixture of complex sinusoids from a
finite number of samples. This problem can be found in numerous
applications, such as direction of arrival estimation
in array signal processing~\cite{LiaoChan16}, bearing and range
estimation in radar signal processing~\cite{CarriereMoses92},
speech analysis in acoustic speech signal
processing~\cite{Flanagan13}, and many
others~\cite{StoicaMoses05}. Traditional LSE solutions include Prony's method~\cite{HauerDemeure90},
MUltiple SIgnal Classification (MUSIC)~\cite{Gruber97}, and
Estimation of Signal Parameters via Rotational Invariance Techniques
(ESPRIT)~\cite{RoyKailath89}. In addition, the LSE problem can
be framed as a sparse recovery problem and solved using the compressed sensing approach~\cite{FangWang16}.

In general, samples of continuous signals are obtained via an analog-to-digital converter
(ADC). Conventional ADCs, however, suffer from a fundamental
bottleneck due to their limited dynamic range (DR). When the
amplitude of the input signal exceeds the DR of a conventional
ADC, clipping/saturation occurs, resulting in distortion. In
such cases, traditional LSE methods experience
considerable performance degradation or may even fail. To avoid
clipping and saturation, automatic gain control is usually
employed at the receiver to maintain a constant output signal
level. Nevertheless, due to the limited DR, a weak signal could be
buried beneath the quantization noise in the presence of a strong
signal.

%via a modulo operation

%In~\cite{BhandariKrahmer20}, it has pointed out that the original
%sample can be uniquely decomposed into its corresponding modulo
%sample and an integer multiple of $2\lambda$. Based on such a
%fact, the noisy modulo sample is modeled as a Gaussian
%mixtures~\cite{MusaJung18} and the generalized approximate message
%passing method is utilized to recover a sparse vector from modulo
%samples. A similar problem (but in noise-free setup) is considered
%in~\cite{PrasannaSriram20}, where the unknown integer multiple of
%$2\lambda$ and the sparse vector are jointly estimated via solving
%a mixed integer linear program. These two works, however, are
%suitable for scenarios where the folding number is relatively
%small.

Recently, unlimited sensing was proposed to address the limited DR
issue of conventional sampling
systems~\cite{BhandariKrahmer17,BhandariKrahmer20}. In an unlimited sensing
framework, a modulo ADC is employed to
map the input signal to a bounded interval
$[-\lambda,\lambda]$ before quantization. Such an operation is
mathematically equivalent to taking the modulo of the input signal
with respect to $\lambda$, which  eliminates the saturation
problem of conventional sampling systems. However, the
modulo operation, which is a nonlinear mapping from
the input to the output, introduces a different type of
information loss. Therefore, how to recover the original input
signal or infer the parameters of interest from modulo samples is
crucial to realizing the potential of the unlimited sensing approach.

The above problem has been extensively investigated over the past few
years. In~\cite{BhandariKrahmer20}, it was shown that, for a
bandlimited signal with a maximum frequency of $\omega$, a
sufficient condition for the recovery of the input signal from its
modulo samples is that the sampling interval should be no greater
than $1/(2\omega e)$, where $e$ is Euler's number. Provided such a
sampling condition is met, there exists a minimal order of
difference, such that the higher-order difference (HOD) of the
original signal is equivalent to the modulo of the
HOD of the modulo samples. Based on this relationship, the
original signal can be recovered (up to a scalar ambiguity) via a higher-order anti-difference operation. This theoretical result represents a breakthrough for data acquisition in
the unlimited sensing framework, and has inspired many
applications in sparse signal recovery~\cite{BhandariKrahmer18-2}, sinusoidal estimation~\cite{BhandariKrahmer18}, state estimation~\cite{WangZheng24} and
computational array signal
processing~\cite{FernandezSamuel21,FernandezSamuel22}. In
addition, unlimited sensing with non-ideal modulo samples~\cite{BhandariKrahmer21},
modulo samples with
hysteresis~\cite{FlorescuKrahmer21,FlorescuKrahmer22}, as well as
non-uniformly sampled modulo samples~\cite{FlorescuKrahmer21-2}
were also studied. Notably, hardware
realization of modulo ADC was presented
in~\cite{BhandariKrahmer21} to validate the performance of
unlimited sensing with non-ideal circuits. The HOD-based methods, however, are very
sensitive to noise. The reason is that the HOD operation shrinks the signal, whereas it has an accumulative
effect on the noise, as the noise has a much wider bandwidth than
the signal. As a result, even with a moderate order
of difference, the effective signal-to-noise ratio (SNR) degrades
substantially, thus leading to a deteriorated performance.

In this paper, we explore the properties of the first-order difference of modulo samples and develop two noise-robust LSE algorithms. Specifically, we
prove that, with a sufficiently high sampling rate, the
first-order difference of the original signal samples equals that
of the modulo samples with a high probability. Based on this
theoretical result, our first proposed algorithm recasts LSE as a robust sparse signal recovery problem. Note that the sparse property was also observed in~\cite{BhandariKrahmer21} and utilized to
recover the original signal. Nevertheless, the
proposed solution~\cite{BhandariKrahmer21} requires the knowledge
of the number of impulsive components as well as the period of the
original signal, which is generally unknown in practice. In \cite{Bhandari22-2}, a
data-driven method was proposed to estimate the number of the
impulsive components. For the second proposed algorithm, we show that, with a sufficiently high sampling rate, the first-order difference of the original signal can be decomposed as a sum of the first-order difference of the modulo samples and a sequence with each element being either zero of $\pm 2\lambda$. This property enables us to cast the problem into a mixed-integer linear program which can be efficiently solved. Simulation results show that both algorithms are robust against noise
and achieve significant performance improvement over the
HOD-based method.

The remainder of the work is organized as follows. In Section II,
preliminaries about modulo ADC and higher-order difference are
introduced. LSE in the unlimited sensing
framework is formulated in Section III. Section IV provides an
overview of the HOD-based approach to the problem. First-order difference-based methods are then
developed in Section V and Section VI. Simulation results are
presented in Section VII, followed by concluding remarks in
Section VIII.

\section{Preliminaries}
The folding operation can be mathematically expressed as the
following nonlinear mapping
\begin{align}
\mathcal{M} _{\lambda }:f \mapsto 2\lambda \left(\left \langle
\frac{f}{2\lambda } +\frac{1}{2} \right \rangle -\frac{1}{2}\right)
\label{folding_oper}
\end{align}
where $\langle a \rangle \triangleq  a-\left \lfloor a \right
\rfloor $ is the fractional part of $a$ and $\lambda \ge 0$ is the
operation range of the modulo ADC. Clearly, such an operation converts
an arbitrary value $f$ into the interval $[-\lambda,\lambda]$.

Define the first-order difference matrix as $\bm
D^1_M\in\mathbb{R}^{(M-1)\times M}$, which is given by
\begin{align}
\bm D_M^1 = \left(\begin{array}{cccccc}
-1 & 1 & 0 & \cdots & 0 & 0 \\
0 & -1 & 1 & \cdots & 0 & 0 \\
\vdots & \vdots & \vdots & \ddots & \vdots & \vdots \\
0 & 0 & 0 & \cdots & -1 & 1
\end{array}\right).
\end{align}
Mathematically, the $(i,j)$th entry of $\bm D_M^1$ is obtained via
$\delta[j-i-1]-\delta[j-i]$ where $\delta[\cdot]$ is the Kronecker
Delta function. Given a vector $\bm x\in\mathbb{C}^{M}$, its
first-order difference vector $\bm{\tilde x}$ can be obtained as
\begin{align}
\bm{\tilde x} = \bm D_M^1 \bm x \in \mathbb{C}^{ M-1 }
\end{align}
where $\bm {\tilde x}(i) = \bm{x}(i+1)-\bm{x}(i)$. Accordingly, the
$N$th-order ($N>1$) difference matrix $\bm D_M^N$ can be
recursively constructed as
\begin{align}
\bm D_M^N = \bm D_{M-N+1}^1 \bm D_{M}^{N-1}.
\end{align}
The $N$th-order difference vector of $\bm{x}$ can be calculated as
\begin{align}
\bm{\tilde x}^{(N)}=\bm D_M^N \bm x \in\mathbb{C}^{ M-N }.
\end{align}

%the modulo of the input signal $y_t$, i.e.,

\section{Problem Formulation}
Consider the following LSE problem
\begin{align}
y_m = \sum_{k=1}^K \alpha_k e^{-j\omega_k m\Delta T} \label{md1}
\end{align}
where $m$ is the sampling index, $\Delta T$ is the
sampling interval, $\omega_k\in[0,2\pi)$ and $\alpha_k$ are,
respectively, the frequency and the complex amplitude of the $k$th
component. The problem of interest is to obtain an estimate of
$\{\alpha_k,\omega_k\}$ from measurements $\{y_m\}_{m=1}^M$. In
practical sampling systems, due to the limited DR of conventional
ADCs, some of the measurements $\{y_m\}$ may be clipped, leading
to severe information loss. To address this difficulty, this work
considers the LSE problem with modulo ADCs. For modulo ADCs, when
the input signal magnitude exceeds a certain threshold $\lambda$,
it will reset such that the signal is folded back to the range
$[-\lambda,\lambda]$. Mathematically, it is equivalent to taking
the remainder of the division of $y_m$ by $\lambda$:
\begin{align}
z_m = \mathcal{U}_{\lambda}(y_m) + v_m \label{md1-2}
\end{align}
where $v_m\sim\mathcal{CN}(0,\sigma^2)$ is the complex additive
white Gaussian noise, and $\mathcal{U}_{\lambda}$ is the modulo
operation performed on a complex value which is defined as
\begin{align}
\mathcal{U}_{\lambda}(a) \triangleq
\mathcal{M}_{\lambda}\big(\text{Re}(a)\big) +
j\mathcal{M}_{\lambda}\big(\text{Im}(a)\big)
\end{align}
in which $\text{Re}(a)$ and $\text{Im}(a)$ respectively denote the
real and imaginary parts of a complex number $a$.

To estimate $\{\omega_k,\alpha_k\}$, we discretize the continuous
frequency parameter space into a finite set of grid points, say
$P\ (P\gg K) $ points in total. Define $\bm a_p \triangleq
[e^{-j\omega_p\Delta T}\ \cdots\ e^{-j\omega_pM\Delta T}]^T$, $\bm A \triangleq
[\bm a_1\phantom{0} \cdots\phantom{0} \bm a_P]$, $\bm \alpha
\triangleq [\alpha_1\ \cdots\ \alpha_P]^T$, $\bm y\triangleq [y_1\
\cdots\ y_M]^T$, $\bm z\triangleq [z_1\ \cdots\ z_M]^T$, and $\bm
v\triangleq [v_1\ \cdots\ v_M]^T$. We can then rewrite~\eqref{md1-2}
in a matrix form as
\begin{align}
\bm z & = \mathcal{U}_{\lambda}(\bm y) + \bm v\notag\\
& = \mathcal{U}_{\lambda}(\bm A\bm \alpha) + \bm v \label{md2}
\end{align}
where $\bm \alpha$ is a $K$-sparse vector. In this work, we assume
that the unknown frequency components lie on the discretized grid.

Now the problem becomes estimating the sparse vector $\bm \alpha$
from $\bm z$. Such a problem can be formulated as
\begin{align}
\min_{\bm {  \alpha}}& \quad \|\bm {  \alpha}\|_0 \notag \\
\text{s.t.}&\quad \|\bm{z}-\mathcal{U}_{\lambda} (\bm{  A}\bm{
\alpha} ) \|_2\le \epsilon \label{opt_1}
\end{align}
where $\|\cdot\|_0$ is the $\ell_0$ norm which counts the number
of nonzero components, and $\epsilon$ is a user-defined error
tolerance. Directly solving~\eqref{opt_1} is intractable since it
involves nonlinear constraints caused by the modulo operation.

We note that~\cite{BhandariKrahmer18} also considered
the LSE problem with modulo samples. The main difference between
our work and~\cite{BhandariKrahmer18} is that we only utilize the
first-order difference of modulo samples, whereas the work
\cite{BhandariKrahmer18} depends on the HOD of the modulo samples
to extract the spectral parameters. In~\cite{BhandariKrahmer18},
the unknown frequency components are not required to lie on the
discretized grid. By resorting to gridless or off-grid compressed
sensing techniques, our proposed method can also be extended to
the scenario where the frequencies do not lie on the
pre-specific discretized grid. In addition, by setting $\omega_k =
k \omega_0$, our signal model has a similar form to that
in~\cite{BhandariKrahmer21}.

\section{Higher-Order Difference-Based Approach}
In this section, we introduce a HOD-based
approach to address the LSE problem.
In~\cite{BhandariKrahmer20}, it was shown that when the sampling
rate is sufficiently high, the HOD (greater than a
certain order) of the original signal $\boldsymbol{y}$
can be obtained from the HOD of modulo samples
$\boldsymbol{z}$. Based on this result, the original signal can be
recovered up to an unknown constant via an anti-difference
operation.

%and $\lambda>0$ is a user-defined threshold parameter

Define $\omega\triangleq \max\{\omega_1,\cdots,\omega_K\}$ and
$B\triangleq \|\bm y\|_{\infty}$. The main result in
\cite{BhandariKrahmer20} is summarized as follows.
\begin{theorem}
\label{t0} Let $y_m$ be samples of $y(t)\in\mathcal{B}_{\omega}$
obtained with a sampling interval of $\Delta T$, where
$\mathcal{B}_{\omega}$ represents the $\omega$-bandlimited
function set. Let $e$ denote Euler's number and
\begin{align}
\bm z = \mathcal{U}_{\lambda}(\bm y)
\end{align}
denote the noise-free modulo sample vector. If the sampling interval $\Delta T$ satisfies
\begin{align}
\Delta T &\le \frac{1}{2 \omega e}\label{cond1}
\end{align}
and meanwhile the order of difference $N$ is no smaller than
\begin{align}
 N &\ge \left\lceil
\frac{\log{\lambda} - \log{B}}{\log{\left(\Delta T\omega
e\right)}}\right\rceil\label{cond2}
\end{align}
then the $N$th-order difference of the original signal $\bm y$ can
be obtained from the $N$th-order difference of its modulo samples
via the following relationship:
\begin{align}
\bm D_M^N \bm{ y} = \mathcal{U}_{\lambda}(\bm D_M^N \bm{ z})
\label{rel_1}
\end{align}
where $\bm D_M^N$ is the $N$th-order difference matrix.
\end{theorem}

\begin{remark}
For finite-length sequences, besides the above requirement on the sampling rate, the number of samples should also satisfy a certain condition in order
to ensure that Theorem 1 holds~\cite{BhandariKrahmer18-2}. Specifically, based on the result
reported in~\cite{BhandariKrahmer18-2}, to recover sinusoidal
mixtures, the required number of modulo samples should be no less
than $2K + 7B/\lambda$. This requirement is mild and can usually
be met in practice.
\end{remark}

Utilizing the above theorem and ignoring the noise, the
optimization~\eqref{opt_1} can be reformulated into a conventional
compressed sensing problem:
\begin{align}
\min_{\bm { \alpha}}& \quad \|\bm { \alpha}\|_0 \notag \\
\text{s.t.}&\quad \mathcal{U}_{\lambda}(\bm D^N_M \bm{z}) =\bm
D^N_M \bm{ A}\bm{ \alpha} \label{opt_2}
\end{align}
which can be efficiently solved by off-the-shelf compressed
sensing solutions, e.g., the orthogonal matching pursuit method~\cite{TroppGilbert07}.
In practice, the modulo samples $\boldsymbol{z}$ are
inevitably corrupted by noise due to quantization errors and
thermal noise. In this case, we have
\begin{align}
\mathcal{U}_{\lambda}(\bm D^N_M \bm{z}) =\bm D^N_M \bm{ A}\bm{
\alpha} + \bm{w} \label{rel_h2}
\end{align}
where $\bm{w}$ represents the error induced by the measurement
noise $\bm v$. The problem is that the HOD operation has a
shrinkage effect on the signal, whereas it has an accumulative
effect on the noise as the noise has a much wider bandwidth than
the signal. As a result, even with a moderate vale of $N$, say
$N=3$, the effective signal-to-noise ratio (SNR) degrades
substantially, thus leading to a deteriorated recovery
performance.

%and $y_m\triangleq y(m\Delta T)$ with $\Delta T$ being the
%sampling interval

\begin{figure}[!t]
\centering
\subfloat[$N=2$]{{\includegraphics[width=0.46\textwidth]{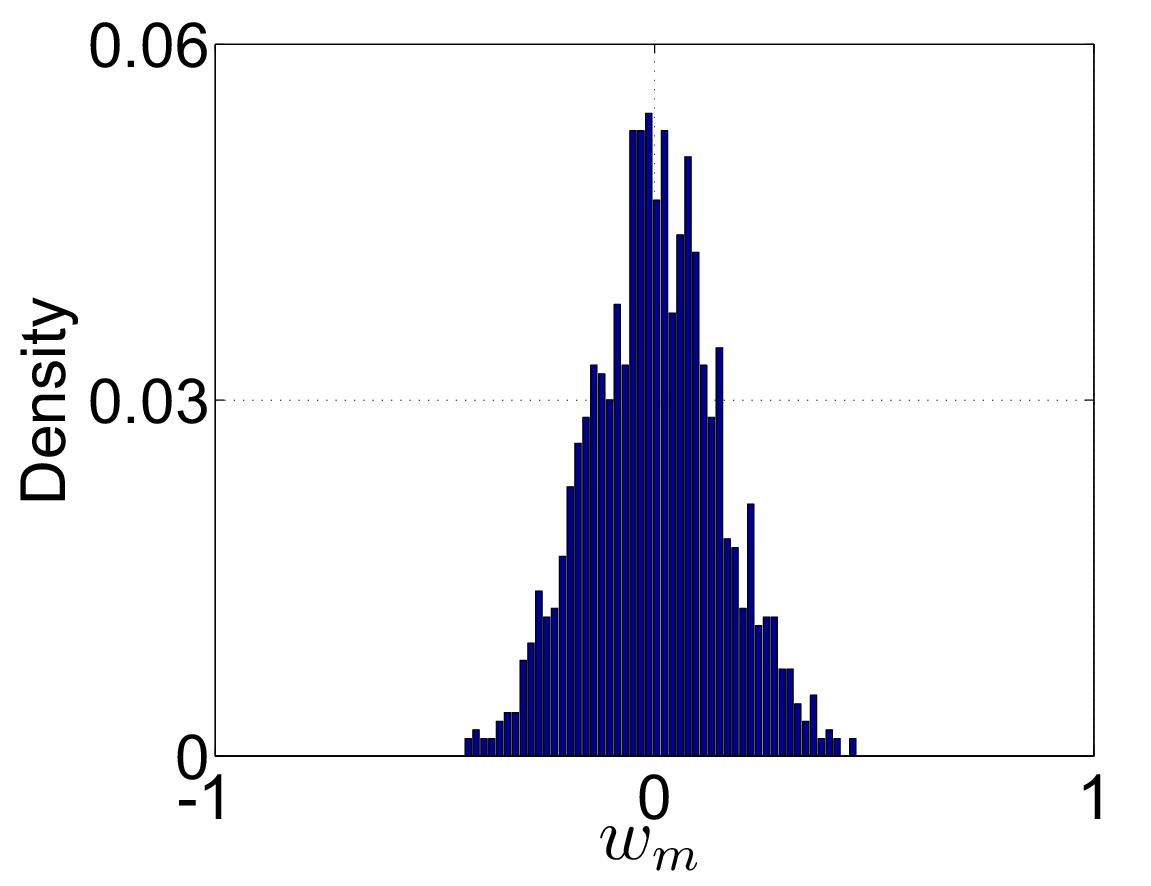} }}\
\subfloat[$N=3$]{{\includegraphics[width=0.46\textwidth]{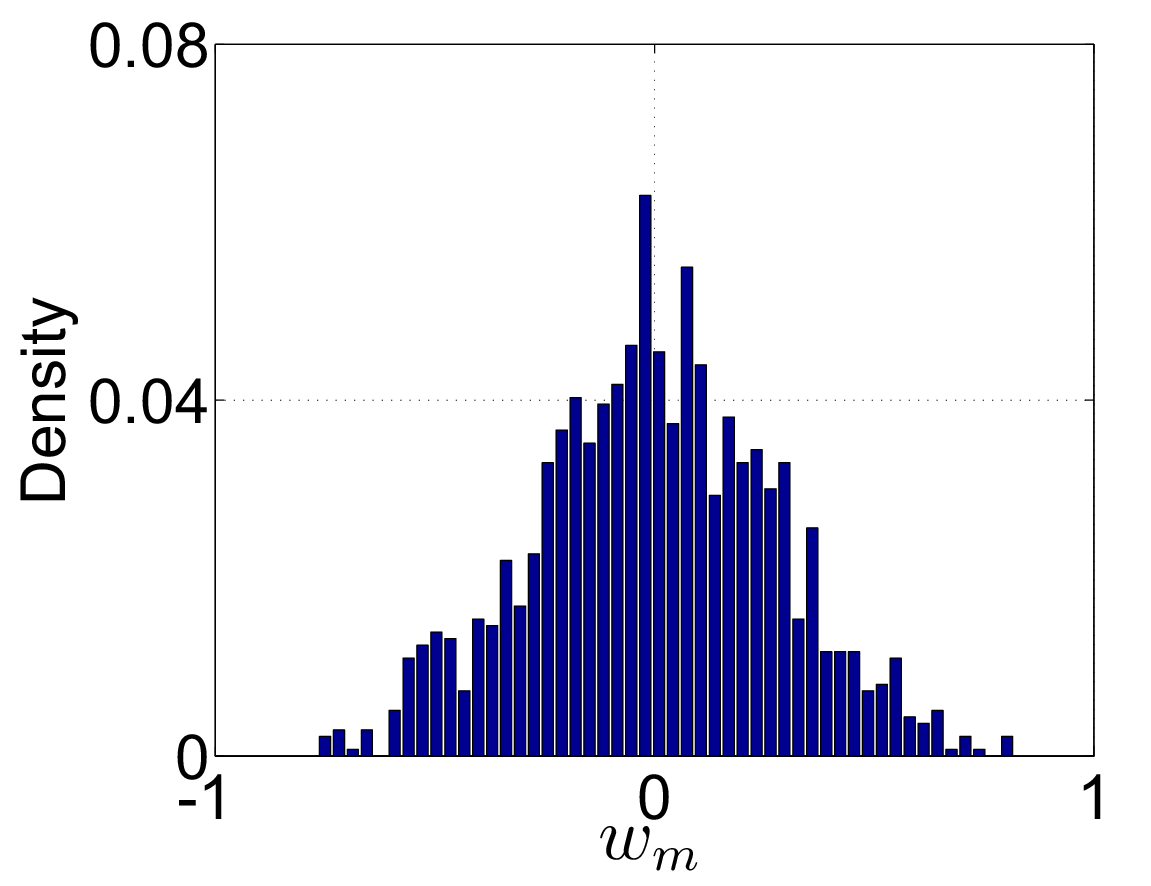} }}\\
\subfloat[$N=4$]{{\includegraphics[width=0.46\textwidth]{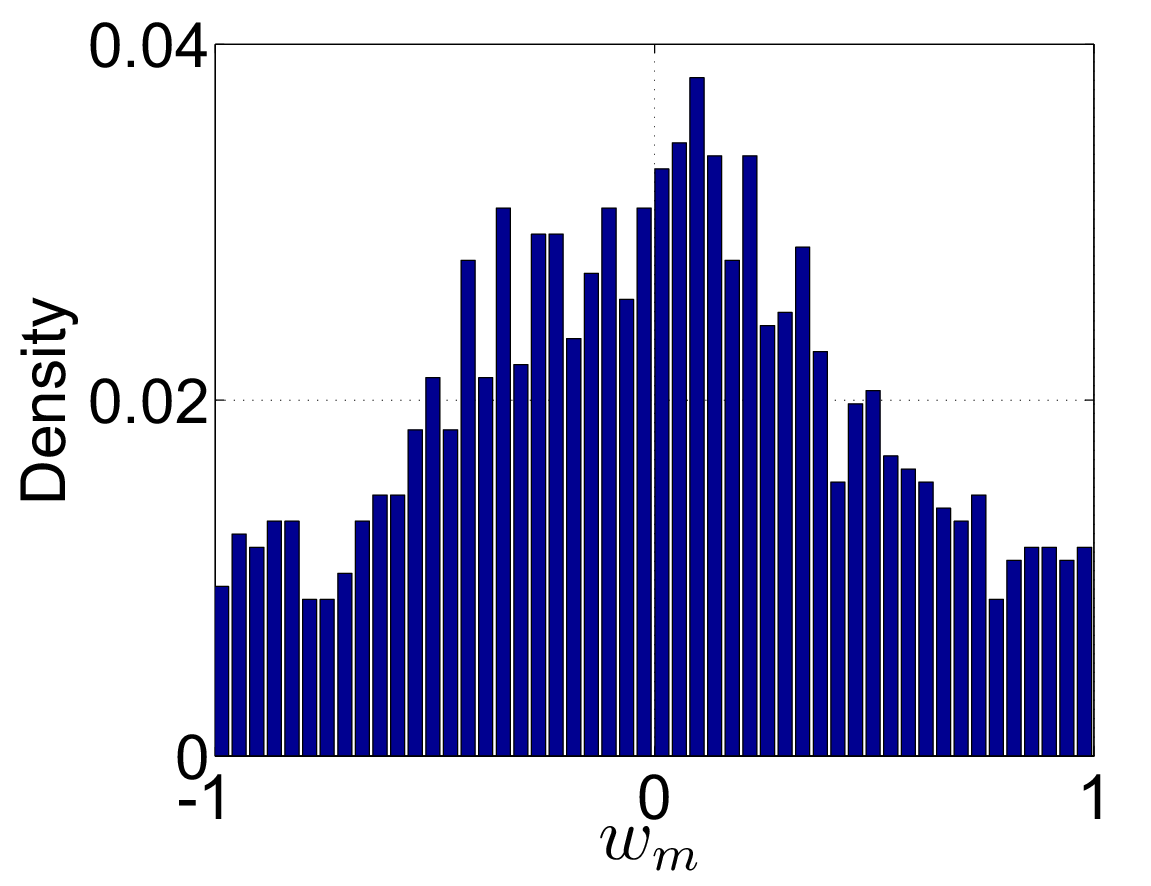} }}\
\subfloat[$N=5$]{{\includegraphics[width=0.46\textwidth]{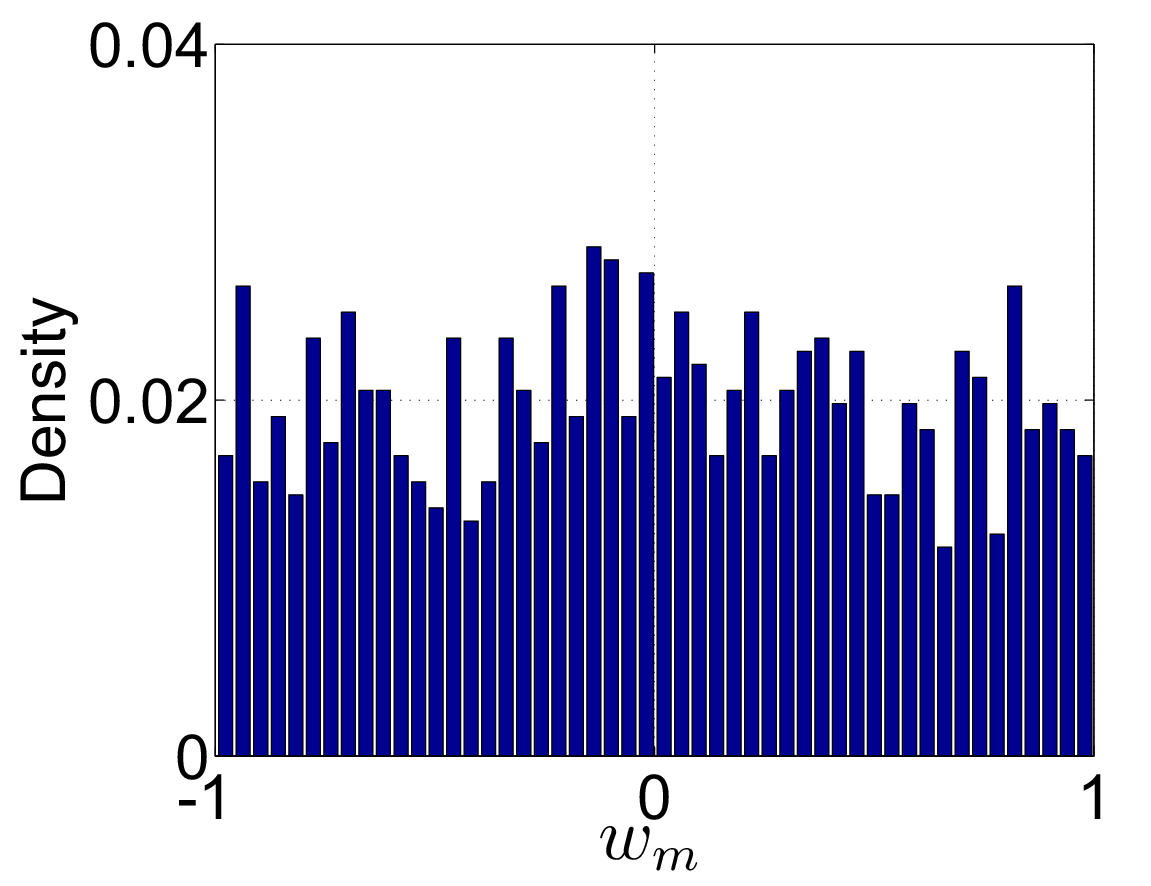} }}
\caption{The histogram of $\{w_m\}$ under different $N$}
\label{f0}
\end{figure}

For demonstration, we consider a
simple example
\begin{align}
y(t) &= 7\sin(1.6\pi t)\\
z_m &=\mathcal{U}_{\lambda}(y_m) + v_m
\end{align}
where $\lambda =1$. In our experiments, we set the sampling
interval $\Delta T = 0.0362$s such that the required conditions in
Theorem~\ref{t0} are satisfied. Fig.~\ref{f0} shows the histogram
of $w_m$ (i.e., the components in $\bm w$) under different choices
of $N$ when the SNR is set to $20$dB. We see that the error $\bm
w$ induced by the measurement noise becomes significantly larger
as $N$ increases.

\begin{comment}
\begin{figure}[!t]
\centering
\includegraphics[width=0.45\textwidth]{err_hist.eps}
\caption{The histogram of $\bm w$ under different $N$}
\label{f0}
\end{figure}
\end{comment}

To overcome this difficulty caused by the
higher-order difference, in the following, we explore the
properties of the first-order difference of modulo samples, and
develop algorithms that are robust against noise.

%noise $\bm{\tilde v}$ could reach the same level as that of the
%signal component $\bm D^N_M \bm y$,
%which will result the effective SNR, defined as $\mathbb{E}(\bm
%D^N_M \bm y)/\mathbb{E}(\bm{\tilde v})$, is even smaller than the
%actual SNR related to the modulo samples $\bm z$. To circumvent
%this issue, we develop two algorithms which only depend on the
%first order difference in the following sections.

\begin{figure}[!t]
\centering \subfloat[The original undistorted samples, samples
obtained via the conventional ADC and modulo samples obtained via
the modulo ADC]
{\includegraphics[width=0.6\textwidth]{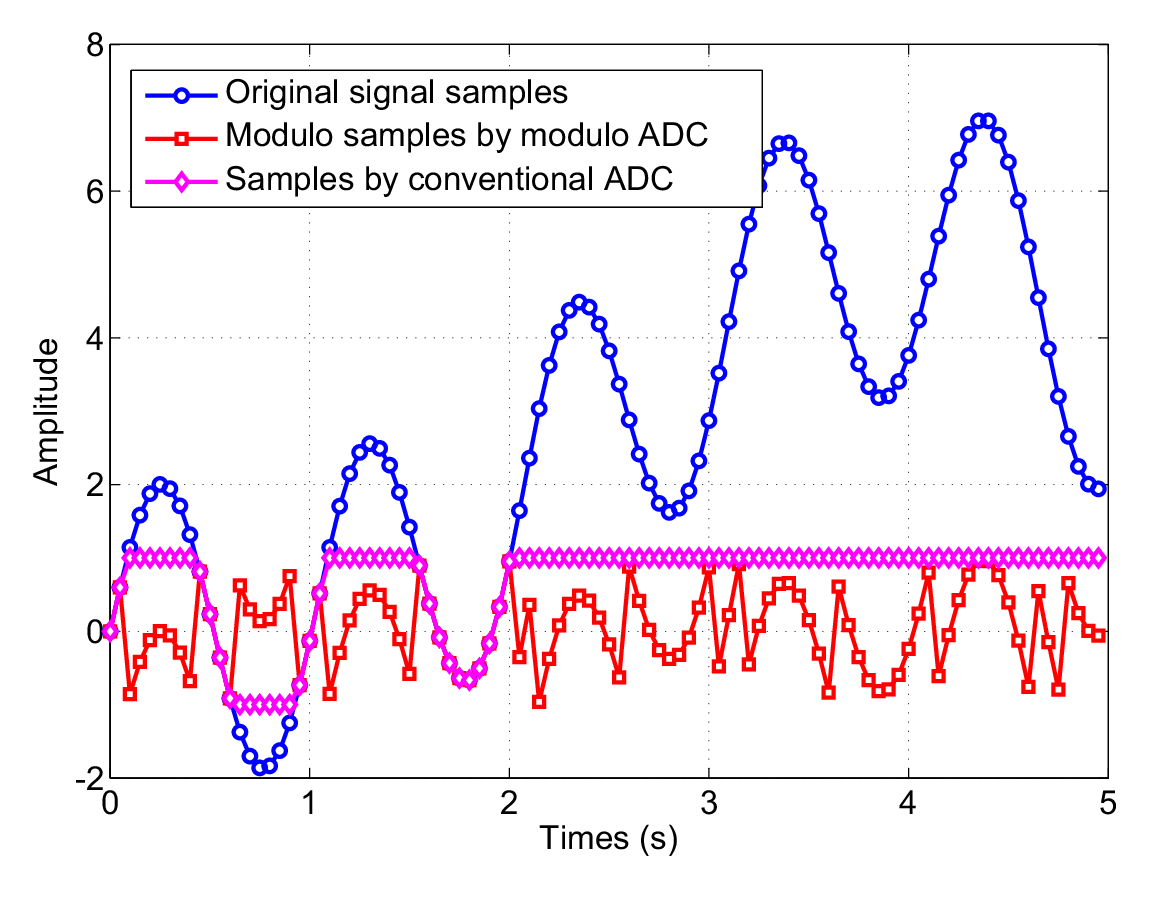}\label{f1a}}\\
\subfloat[The-first order differences of the original samples and
modulo samples]
{\includegraphics[width=0.6\textwidth]{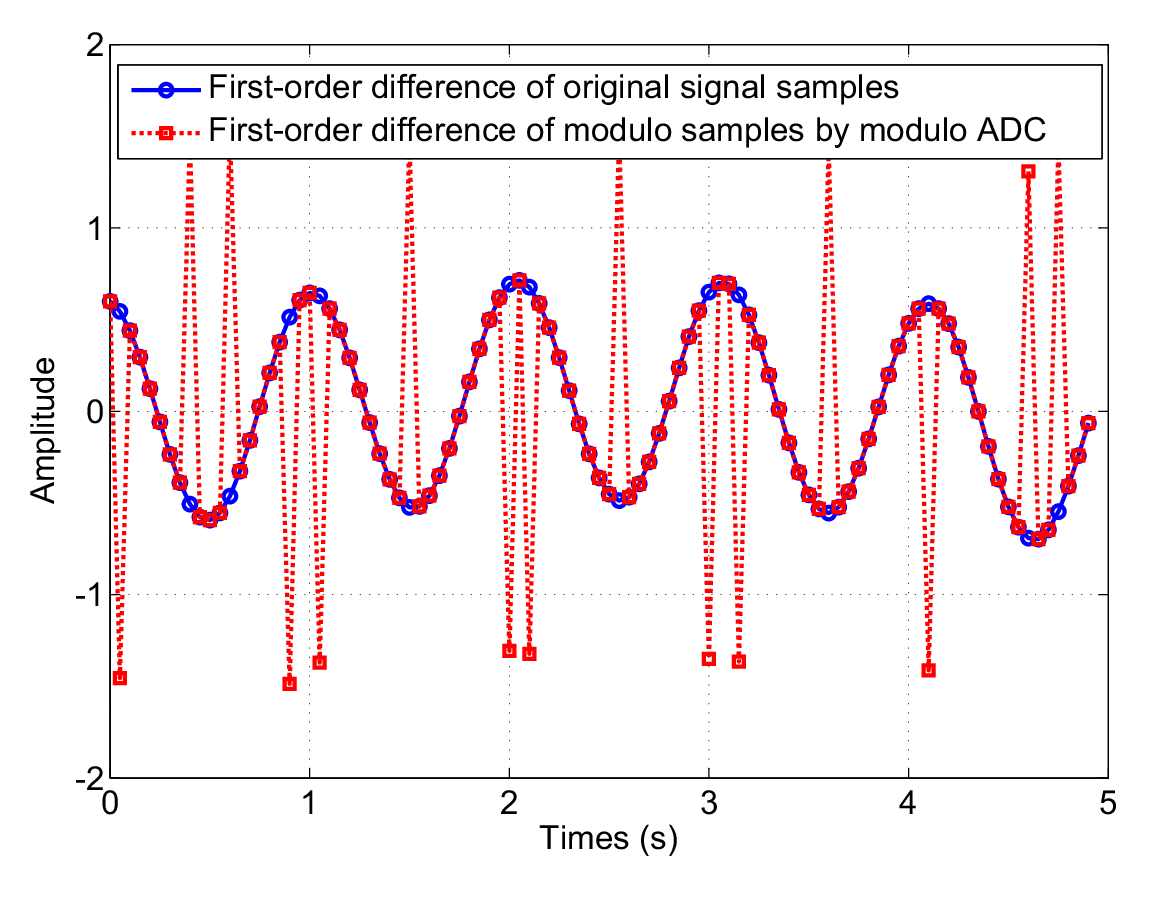}\label{f2a}}
\caption{First-order differences of original samples and modulo
samples.} \label{f1}
\end{figure}

\section{Robust Compressive Sensing Based Approach}
\label{sec_rcs}
%\subsection{Motivations}
This proposed approach is based on the observation that, although
the first-order difference of the modulo samples are not exactly the
same as the first-order difference of the original signal samples,
these two indeed coincide with a high probability
when the sampling rate is sufficiently high. To illustrate this,
consider the following example
\begin{align}
y(t) = 4\sin(\pi t/6)-2 \sin(\pi t/3) +  2\sin(2\pi t- 0.2t)
\end{align}
Setting the sampling interval to $0.05$ seconds, Fig.~\ref{f1}(a)
shows the undistorted original samples, the samples obtained by a
conventional ADC, and the modulo samples obtained by a modulo ADC. The
measurable ranges of the conventional ADC and modulo ADC are set to
$[-0.5,0.5]$. In Fig.~\ref{f1}(b), the first-order difference of
the original samples and that of the modulo samples are plotted.
We see that most of the entries are identical. An intuitive
explanation of this fact is that when the sampling rate is
high, the original signal
undergo slow variations between successive samples. As a result, the
first-order difference of the original signal samples equals to
that of the modulo samples with a high probability.

In fact, those inconsistent samples can be considered as outliers.
Thus the LSE problem can be reformulated as a
robust sparse signal recovery problem. First, we formulate the
first-order difference of modulo samples as
\begin{align}
\bm{\tilde z}&\triangleq \bm D_M^1 \bm z\notag\\
& = \bm D_M^1 \mathcal{U}_{\lambda}(\bm A\bm \alpha) + \bm D_M^1 \bm v\notag\\
& = \bm D_M^1\bm A\bm \alpha +\bm s +  \bm D_M^1 \bm v\notag\\
& \triangleq \bm {\tilde  A} \bm \alpha +\bm {\tilde v} + \bm s
\label{md3}
\end{align}
where $\bm {\tilde A}\triangleq \bm D_M^1\bm A$, $\bm {\tilde v} \triangleq
\bm D_M^1 \bm{ v}$, and $\bm s$ is an unknown, sparse outlier
vector. Note that the above equation comes from the fact that when
the sampling rate is sufficiently high, the first-order difference
of the modulo samples can be expressed as a sum of the first-order
difference of original signal samples and a sparse outlier vector,
i.e.
\begin{align}
\bm D_M^1 \mathcal{U}_{\lambda}(\bm A\bm \alpha)=\bm D_M^1\bm A\bm
\alpha +\bm s
\end{align}

The problem of estimating $\bm \alpha$ from~\eqref{md3} is a
classic robust sparse signal recovery problem which has been
extensively investigated, e.g.~\cite{WrightYang08,NguyenTran12,
NguyenTran13,CarrilloBarner13,OllilaKim14,ParkerSchniter14,WanDuan17}.
Specifically, a theoretical analysis conducted
in~\cite{NguyenTran12} shows that under certain conditions,
exact recovery of the sparse vector $\bm \alpha$ from sparse
outlier-corrupted measurements can be guaranteed. In addition to
theoretical justifications, many robust compressed sensing
algorithms were developed from either optimization-based or
Bayesian inference frameworks. Popular algorithms include the
outlier-compensation-based
approach~\cite{CarrilloBarner13,OllilaKim14}, and outlier
identify-and-reject approach~\cite{WanDuan17}.
Algorithm~\ref{algo1} summarizes the proposed robust
compressive sensing (RCS) based solution for the LSE problem with
modulo samples.

\begin{algorithm}[!t]
  \caption{RCS-based algorithm for LSE problem with modulo samples} \label{algo1}
  \begin{algorithmic}[0]
    \State \textbf{Input:} $\bm z$ and $M$.
    \State \textbf{Output:} $\bm \alpha$ and the reconstructed original signal $\bm {\hat y}$.
  \State 1)\ Construct $\bm A$ and $\bm D_M^1$;
  \State 2)\ Compute $\bm{\tilde A} = \bm D_M^1\bm A$ and $\bm{\tilde z}=\bm D_M^1\bm z$;
  \State 3)\ Estimate $\bm \alpha$ via a robust compressive
  sensing method (e.g.,~\cite{WanDuan17});
  \State 4)\ Reconstruct $\bm{\hat y} = \bm A \bm \alpha$.
  \end{algorithmic}
\end{algorithm}

Clearly, exact recovery of the sparse signal $\bm \alpha$ is
feasible when most measurements of $\bm{\tilde z}$ are not
outliers, in other words, when $\bm s$ is sufficiently sparse.
Intuitively, given other parameters fixed, the
sparsity of the outlier vector $\bm s$ is related to the sampling
rate: the higher the sampling rate, the sparser the outlier
vector. In the following, we provide a quantitative analysis which
establishes a mathematical relationship between the sampling rate
and the sparsity of the outlier vector $\bm s$ obtained via the
first-order difference operation.

\begin{remark}
One issue with the compressive sensing approach is whether the sensing matrix satisfies the restricted isometry property (RIP). While the RIP is generally intractable to verify for generic dictionaries, recent
studies showed that the isometry constant $\delta_K$ can be upper
bounded by the mutual coherence $\mu$ of the matrix as follows~\cite{Elad07,Yang14}
\begin{align}
\delta_K\leq (K-1)\mu
\end{align}
Therefore, we can check the RIP of $\bm{\tilde A}$ by examining
its mutual coherence. Taking $M = 400$ and $P=20$ as an example (the same setup as in our simulations), we find that when $\Delta T\ge 0.045$, the mutual coherence $\mu$ of $\bm{\tilde A}$ is consistently less than $0.1$. Under such a circumstance, we can
conclude that
\begin{align}
\delta_K \le 0.1(K-1)
\end{align}
To guarantee $\delta_{2K} < 1$, one has $K < 5.5$, meaning that
$\bm{\tilde A}$ satisfies the RIP condition when $K $ is no
greater than $5$.
\end{remark}

\subsection{Theoretical Analysis}
The line spectral model in~\eqref{md1} can be re-expressed as
\begin{align}
y_m &= \sum_{k=1}^K |\alpha_k|\left(\cos(\omega_k m\Delta T-\phi_k) -
j\sin(\omega_k m\Delta T-\phi_k)\right)\notag\\
&\triangleq f_m^r - j f_m^i
\end{align}
where $\phi_k = \angle{\alpha_k}$ denotes the argument of the
complex parameter $\alpha_k$, $f_m^r$ and $f_m^i$ are,
respectively, the real and imaginary parts of $y_m$ which are
given as
\begin{align}
f_m^r &= \sum_{k=1}^K |\alpha_k|\cos(\omega_k m\Delta T-\phi_k)\\
f_m^i &= \sum_{k=1}^K |\alpha_k|\sin(\omega_k m\Delta T-\phi_k)\notag\\
& = \sum_{k=1}^K |\alpha_k|\cos(\omega_k m\Delta T-\psi_k)
\end{align}
where $\psi_k = \phi_k +\pi/2 $. We see that the two sinusoidal
mixture functions $f_m^r$ and $f_m^i$ share the same frequency
components with different phases. To analyze the relationship
between the first-order difference of $y_m$ and the first-order
difference of its modulo samples, if suffices to examine the
first-order difference of either its real component or its
imaginary component.

Specifically, consider the following sinusoidal
mixture
\begin{align}
f(t) = \sum_{k=1}^K \beta_k \cos(\omega_k t + \varphi_k)
\label{eqn1}
\end{align}
where $\varphi_k \in (0,2\pi]$, $\omega_k \ge 0$ and $\beta_k
\in\mathbb{R}$ are, respectively, the phase, frequency and
amplitude of the $k$th component. Define $\omega \triangleq
\max\{\omega_1,\cdots,\omega_K\}$, $\bar \beta   \triangleq
\sum_{k = 1}^K |\beta_k \bar \omega_k|$, and $\bar \omega_k =
\omega_k/\omega$. Let $f_m = f(m\Delta T)$ and $x_m =
\mathcal{M}_{\lambda}(f_m)$ respectively denote the original
sample and the modulo sample of $f(t)$, where $\Delta T\in
\mathbb{R}^+$ is the sampling interval. Let $\tilde f_m \triangleq
f_{m+1} - f_{m}$ denote the first-order difference of the original
samples and $\tilde x_m \triangleq x_{m+1} - x_{m}$ denote the
first-order difference of the modulo samples.

According to~\eqref{folding_oper}, the modulo sample of $f_m$ can
be written as
\begin{align}
x_m = 2\lambda\left(\frac{f_m}{2\lambda}-\left\lfloor
\frac{f_m}{2\lambda} +\frac{1}{2}\right\rfloor\right)
\end{align}
The above equation can be re-expressed as
\begin{align}
f_m = 2\lambda \left\lfloor \frac{f_m}{2\lambda} + \frac{1}{2} \right\rfloor + x_m
\label{exp_fm}
\end{align}
Here $f_m$ is a sum of two terms, where the first term is the
quantization output of $f_m$ that is quantized by a step size of
$2\lambda$, and the second term $x_m$ can be considered as the
quantization noise.

%~\cite{Barakat74,BarakatCole79,EditionPapoulis02}

In this paper, we make the following widely-used assumption~\cite{Barakat74,StoicaMoses05,BarakatCole79,EditionPapoulis02}:
\begin{assumption}
The phase of each component in $f(t)$ is a random variable which
follows a uniform distribution, i.e.,
\begin{align}
p(\varphi_k) = \left\{
\begin{array}{ll}
\frac{1}{2\pi},&\varphi_k\in(-\pi,\pi]\\
0,&\text{otherwise}
\end{array}
\right.
\end{align}
\end{assumption}
Under such an assumption, $x_m$ can be modeled as a uniform random
variable when $2\lambda$ is sufficiently small:
\begin{align}
x_m\sim \mathbb{U}(-\lambda,\lambda) \label{model_pm}
\end{align}
A rigorous justification of the uniform distribution of $x_m$ is
discussed in Appendix~\ref{dis}.

We are now in a position to present our main result, which is
summarized as follows.

\begin{theorem}
\label{t1} Consider a sinusoidal mixture signal $f(t)$ which has a
form of~\eqref{eqn1}, where the phase of each sinusoidal component
follows a uniform distribution $\mathbb{U}(-\pi,\pi)$. If the
sampling interval satisfies
\begin{align}
\Delta T \le \frac{1}{2\omega}\left(\frac{\bar
\beta}{2\lambda}\right)^{-1} \label{eqn2}
\end{align}
then the first-order difference of the original samples equals
the first-order difference of the modulo samples with probability
exceeding
\begin{align}
\text{Pr}(\tilde f_m = \tilde x_m) \triangleq  p \ge 1 - \frac{\Delta
T \omega \bar \beta}{2\lambda }
\label{ppr}
\end{align}
\end{theorem}
\begin{IEEEproof}
See Appendix~\ref{appA}.
\end{IEEEproof}

%Notice that the concept of bandlimited CF was first introduced
%in~\cite{WidrowKollar96} to study the statistical properties of
%the quantization noise. As indicated in~\cite{WidrowKollar96},
%although most random variables encountered in practice do not have
%perfectly bandlimited CFs, their CFs are often approximately
%bandlimited.

Theorem~\ref{t1} indicates that the probability of $\tilde f_m =
\tilde x_m$ can be made arbitrarily close to 1 by increasing the
sampling rate $1/\Delta T$. Specifically, to make sure that $p$ is
no less than a pre-specified threshold $\tau$, i.e., $p \ge \tau$,
the sampling interval needs to satisfy
\begin{align}
\Delta T \le (1-\tau) \frac{2\lambda}{\omega\bar
{\beta}}\triangleq \frac{1}{2\omega \eta_1} \label{dt_p}
\end{align}
where $\eta_1$ is defined as
\begin{align}
\eta_1 \triangleq \frac{1}{2(1-\tau)}\bar N
\end{align}
with $\bar N \triangleq \bar \beta / (2\lambda)$ denoting the
effective number of folding times. We see that the sampling
condition in~\eqref{dt_p} has a form similar to the condition
in~\eqref{cond1}, except that they have different scaling factors.

%\begin{remark}
%Although the LSE problem is considered in this work, it should be
%noted that our main results (i.e., Theorem~\ref{t1} and latter
%Theorem~\ref{t2}) can be extended to accommodate bandlimited
%signals with continuous spectrum.
%\end{remark}

\section{Mixed-Integer Linear Programming Based Recovery Approach}
\subsection{Motivations and Algorithm Development}
It is known that the original sample $f_m$ and the modulo sample
$x_m$ satisfy the unique decomposition property: $x_m = f_m +
2\lambda e_m$, where $e_m$ is an integer. Similarly, it can be
readily verified that the first-order difference of the original
samples and the first-order difference of the modulo samples are
related as $\tilde x_m = \tilde f_m +2\lambda \tilde e_m$, where
$\tilde{e}_m$ is an integer. In the previous section, we observed and
proved that given a sufficiently high sampling rate,
$\{\tilde {e}_m\}$ is a sparse vector with only few nonzero entries.
Thus the LSE problem can be cast as a
robust sparse signal recovery problem. This formulation,
however, fails to utilize the information that $\tilde {e}_m$ is an
integer.

In fact, considering the property that $e_m$ is an integer, the
LSE problem with noise-free $\bm z$ can be
formulated as a combinatorial
problem~\cite{PrasannaSriram20,MusaJung18}
\begin{align}
\min_{\bm \alpha,\bm e}&\quad \|\bm \alpha\|_1\notag\\
\text{s.t.}&\quad \bm A\bm \alpha = \bm z + 2\lambda \bm e \notag\\
&\quad \text{Re}(\bm e)\in \mathbb{Z}^{M}, \ \text{Im}(\bm e)\in \mathbb{Z}^{M} \label{ori_MILP}
\end{align}
where $\mathbb{Z}^{M}$ represents the $M$-dimensional integer
space. Such an optimization problem can be transformed into a
mixed integer linear program and solved via the branch-and-bound
algorithm~\cite{Clausen99}. This approach, however, is not
suitable when the folding number $\lfloor
f_m/(2\lambda)+1/2\rfloor$ is large as the search space increases
exponentially with the folding number.

To reduce the search space, in this section, we take the
first-order difference of modulo samples. Specifically, we have
\begin{align}
\bm{\tilde z}&\triangleq \bm D_M^1 \bm z \notag\\
&= \bm D_M^1(\mathcal{U}_{\lambda}(\bm A\bm \alpha) + \bm v)\notag\\
& = \bm D_M^1 ( \bm A\bm \alpha + 2\lambda \bm e +  \bm v)\notag\\
& = \bm D_M^1\bm A\bm \alpha +2\lambda \bm D_M^1 \bm e +  \bm D_M^1 \bm v\notag\\
& \triangleq \bm {\tilde  A} \bm \alpha +2\lambda \bm {\tilde e} +
\bm {\tilde v} \label{fod_1}
\end{align}
where $\bm {\tilde e}\triangleq \bm D_M^1 \bm e$ is an unknown
vector with $\text{Re}(\bm {\tilde e})\in \mathbb{Z}^{M-1}$ and
$\text{Im}(\bm {\tilde e})\in \mathbb{Z}^{M-1}$. We will show that
when the sampling rate exceeds a certain threshold, both the real and
imaginary components of entries in $\bm{\tilde e}$ belong to the set
$\{0,\pm 1\}$. Thus we can formulate the LSE problem as
\begin{align}
\min_{\bm \alpha,\bm{\tilde e}}&\quad \|\bm \alpha\|_1\notag\\
\text{s.t.}&\quad \|\bm{\tilde z} - \bm {\tilde A}\bm \alpha + 2\lambda \bm{\tilde e}\|_2^2 \le \epsilon \notag\\
&\quad \text{Re}(\bm{\tilde e})\in \{0,\pm 1\}^{(M-1)}\notag\\
&\quad \text{Im}(\bm{\tilde e})\in \{0,\pm 1\}^{(M-1)}
\label{noisy_MILP}
\end{align}
where $\epsilon$ is a user-defined parameter. For convenience, we
write~\eqref{noisy_MILP} into a real-valued form, i.e.,
\begin{align}
\min_{\bm {\check \alpha},\bm{\check e}}&\quad \|\bm {\check \alpha}\|_1\notag\\
\text{s.t.}&\quad \|\bm{\check z} - \bm {\check A}\bm {\check \alpha} +
2\lambda \bm{\check e}\|_2^2 \le \epsilon \notag\\
&\quad \bm{\check e}\in \{0,\pm 1\}^{2(M-1)}
\label{real_noisy_MILP}
\end{align}
where $\bm{\check \alpha} \triangleq [\text{Re}(\bm {\alpha})^T\ \text{Im}(\bm {\alpha})^T]^T$, $\bm{\check e} \triangleq [\text{Re}(\bm {\tilde e})^T\ \text{Im}(\bm {\tilde e})^T]^T$, $\bm{\check z} \triangleq [\text{Re}(\bm {\tilde z})^T\ \text{Im}(\bm {\tilde z})^T]^T$, and
\begin{align*}
\bm {\check A} \triangleq \begin{bmatrix} \text{Re}(\bm {\tilde
A})&-\text{Im}(\bm {\tilde A})\\\text{Im}(\bm {\tilde
A})&\text{Re}(\bm {\tilde A}) \end{bmatrix}
\end{align*}

For the $p$th component of $\bm{\check \alpha}$, i.e., $\check
\alpha_p$, we define two auxiliary variables $\xi_p = \max\{\check
\alpha_p,0\}$ and $\zeta_p = \max\{-\check \alpha_p,0\}$ such that
\begin{align}
\check \alpha_p &= \xi_p - \zeta_p\\
|\check \alpha_p|&= \xi_p + \zeta_p
\end{align}
Therefore we can re-express $\bm{\check \alpha}$ and $\|\bm
{\check \alpha}\|_1$ as
\begin{align}
\bm {\check\alpha} & = \bm \xi - \bm \zeta\\
\|\bm {\check\alpha}\|_1& = \bm 1^T(\bm \xi + \bm \zeta)
\end{align}
where $\bm \xi\in\mathbb{R}_+^{P}$ and $\bm
\zeta\in\mathbb{R}_+^{P}$ are vectors with $\xi_p$ and $\zeta_p$
being their $p$th component respectively, and $\mathbb{R}_+$ denotes the set of non-negative real numbers. Such a representation leads to
\begin{align}
\min_{\bm \xi , \bm \zeta,\bm {\check e}}\quad &\bm 1^T(\bm \xi + \bm \zeta)\notag\\
\text{s.t.}\quad& \|\bm{\check z} - \bm {\check A}\left(\bm \xi - \bm \zeta\right) -
2\lambda \bm{\check e}\|_2^2\le \epsilon\notag\\
&  \bm{\check e}\in\{0,\pm 1\}^{2(M-1)} \notag\\
& \bm \xi \in \mathbb{R}_+^{P}, \ \bm \zeta \in \mathbb{R}_+^{P}
\label{opt1}
\end{align}
Unfortunately,~\eqref{opt1} is a mixed-integer quadratic
optimization problem, which is in general intractable. To
deal with this challenge, we replace the quadratic inequality
constraint by a linear constraint, resulting in the following
optimization problem
\begin{align}
\arg\min_{\bm \xi , \bm \zeta,\bm {\check e}}\quad &\bm 1^T(\bm \xi + \bm \zeta)\notag\\
\text{s.t.}\quad&  - \epsilon' \bm 1  \preceq \left(\bm{\check z}
-\bm {\check A}\left(\bm \xi - \bm \zeta\right) -
2\lambda \bm{\check e}\right) \preceq \epsilon' \bm 1   \notag\\
&  \bm{\check e}\in\{0,\pm 1\}^{2(M-1)} \notag\\
& \bm \xi \in \mathbb{R}_+^{P}, \ \bm \zeta \in \mathbb{R}_+^{P}
\label{opt2}
\end{align}
where $\epsilon'$ is a user-defined parameter and $\preceq$
denotes the element-wise inequality. The problem~\eqref{opt2} is a
mixed-integer linear programming that can be solved via the
branch-and-bound algorithm (e.g., the off-the-shelf tool
{\it{intlinprog}} in Matlab). Once we obtain $\bm \xi$, $\bm
\zeta$, and $\bm {\check e}$, we can construct $\bm{\check \alpha}
= \bm \xi - \bm \zeta$. The mixed-integer linear programming
(MILP) based algorithm for the LSE problem with modulo samples is
summarized in Algorithm~\ref{algo2}.

%With $\bm{\tilde \alpha}$, we can immediately obtain $\bm\alpha$,

%followed by the estimates of $\{\omega_k,\alpha_k\}$.

\begin{algorithm}[!h]
  \caption{MILP-based algorithm for LSE problem with modulo samples} \label{algo2}
  \begin{algorithmic}[0]
    \State \textbf{Input:} $\bm z$, $M$, and the error tolerance $\epsilon$.
    \State \textbf{Output:} $\bm \alpha$ and the reconstructed original signal $\bm {\hat y}$.
  \State 1)\ Construct $\bm A$ and $\bm D_M^1$;
  \State 2)\ Compute $\bm{\tilde A} = \bm D_M^1\bm A$ and $\bm{\tilde z}=\bm D_M^1\bm z$;
  \State 3)\ Estimate $\bm \xi$ and $\bm \zeta$ by solving the optimization problem
  in~\eqref{opt2} (e.g., via {\it{intlinprog}} in Matlab);
  \State 4)\ Calculate $\bm \alpha$ using $\bm \xi$ and $\bm \zeta$;
  \State 5)\ Reconstruct $\bm{\hat y} = \bm A \bm \alpha$.
  \end{algorithmic}
\end{algorithm}

\begin{figure*}[!t]
\centering
{\subfloat {\includegraphics[width=0.32\textwidth]{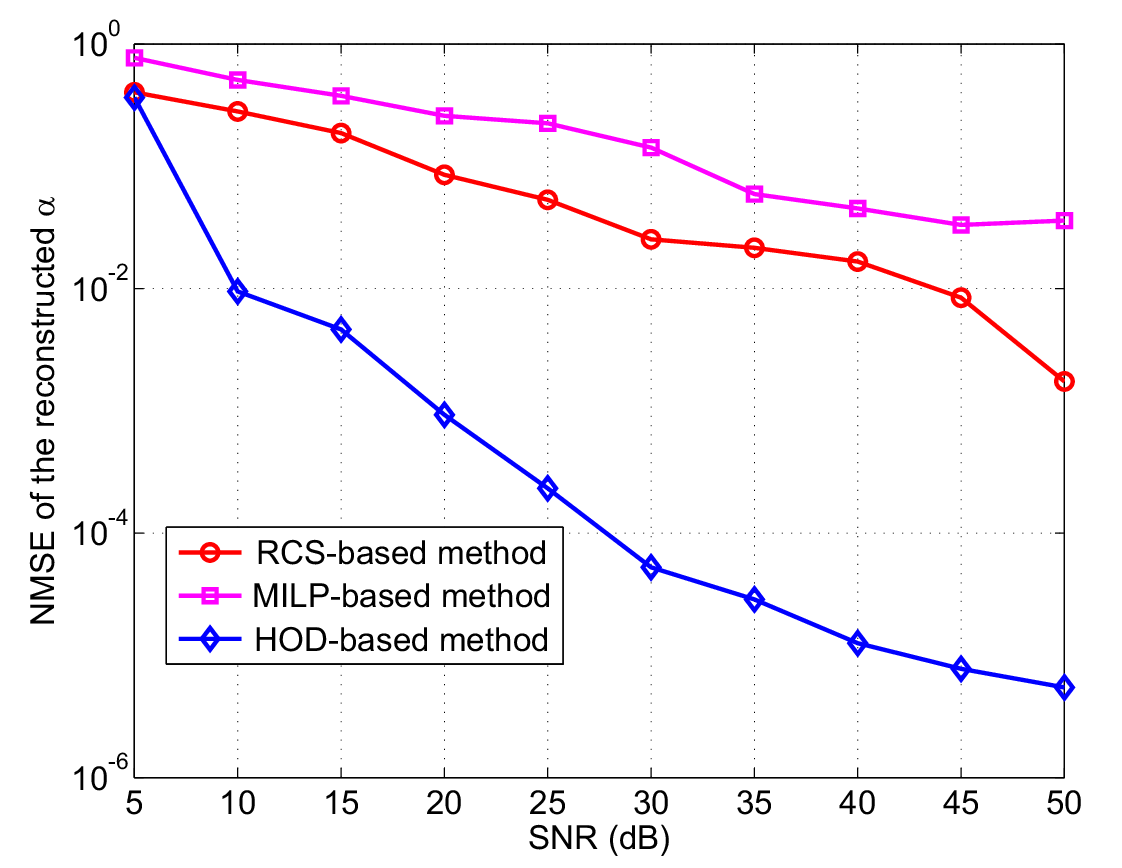}}}
\ \subfloat{\includegraphics[width=0.32\textwidth]{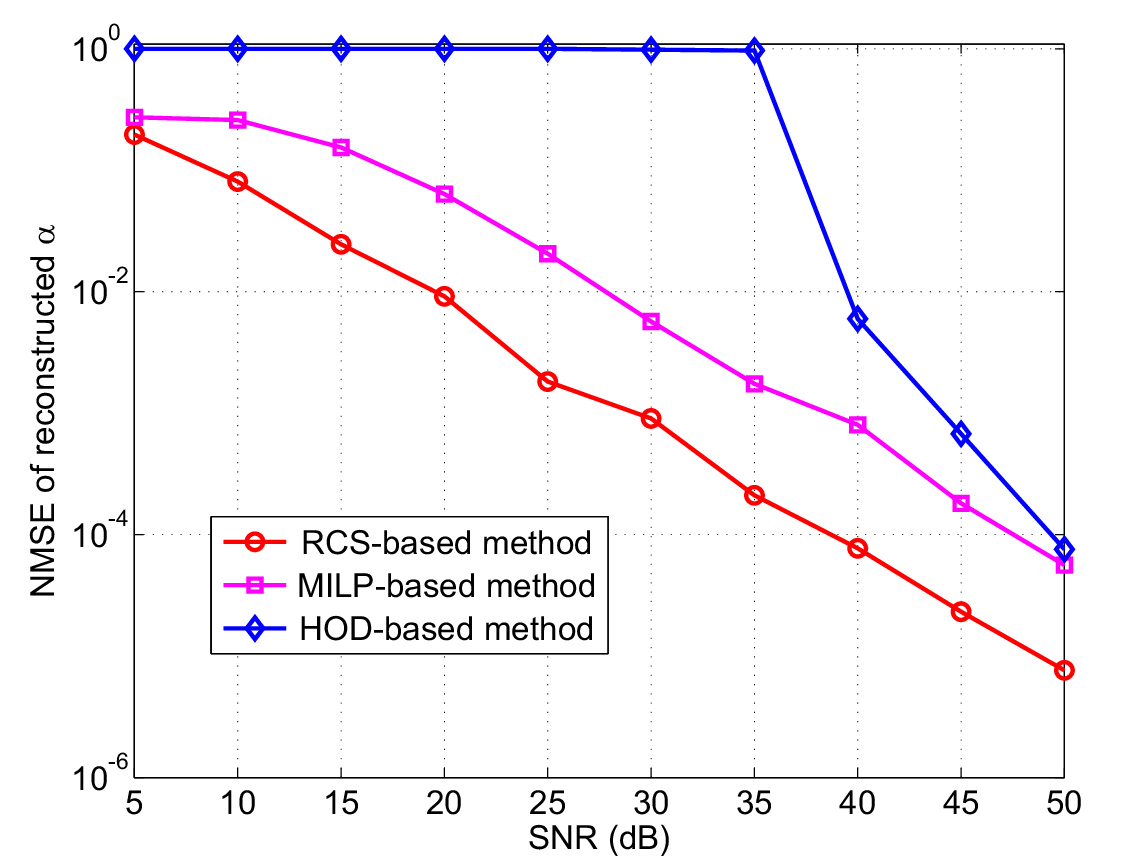}}
\ \subfloat{\includegraphics[width=0.32\textwidth]{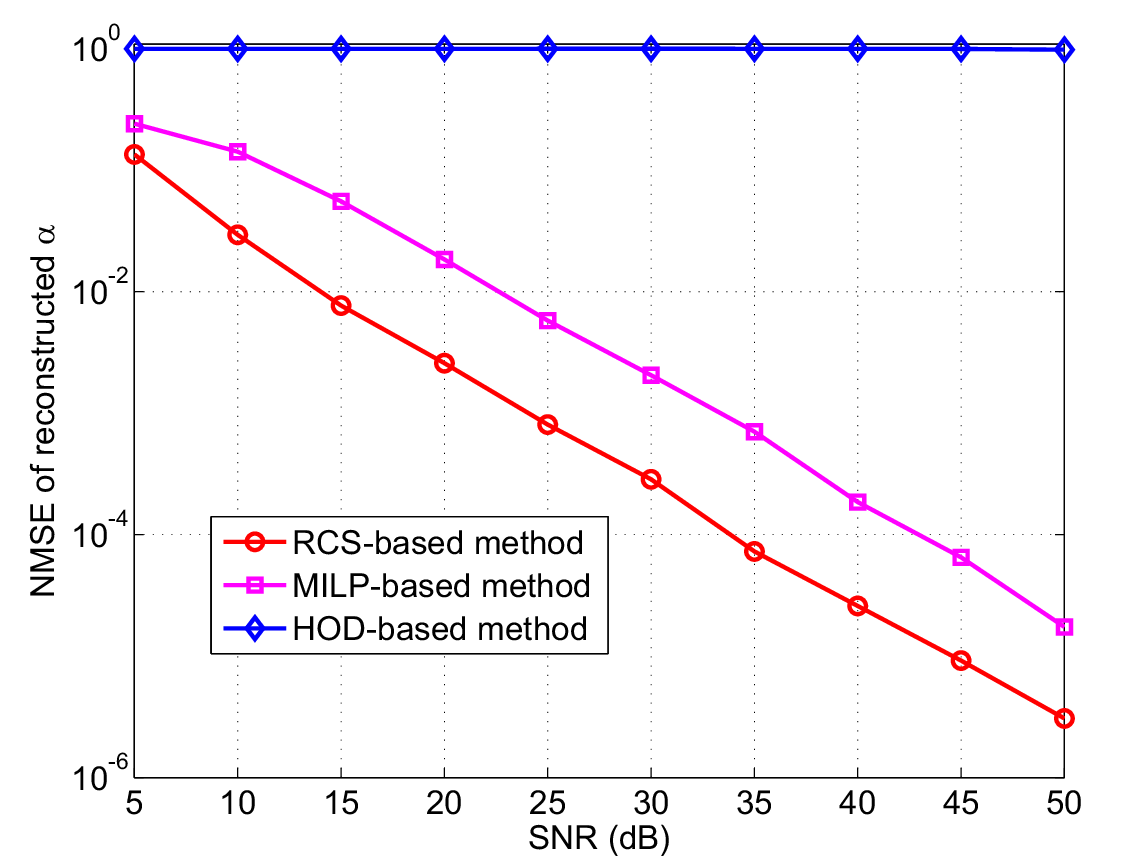}}\\

{\setcounter {subfigure} {0}\subfloat[$\Delta T = 0.004$s]{\includegraphics[width=0.32\textwidth]{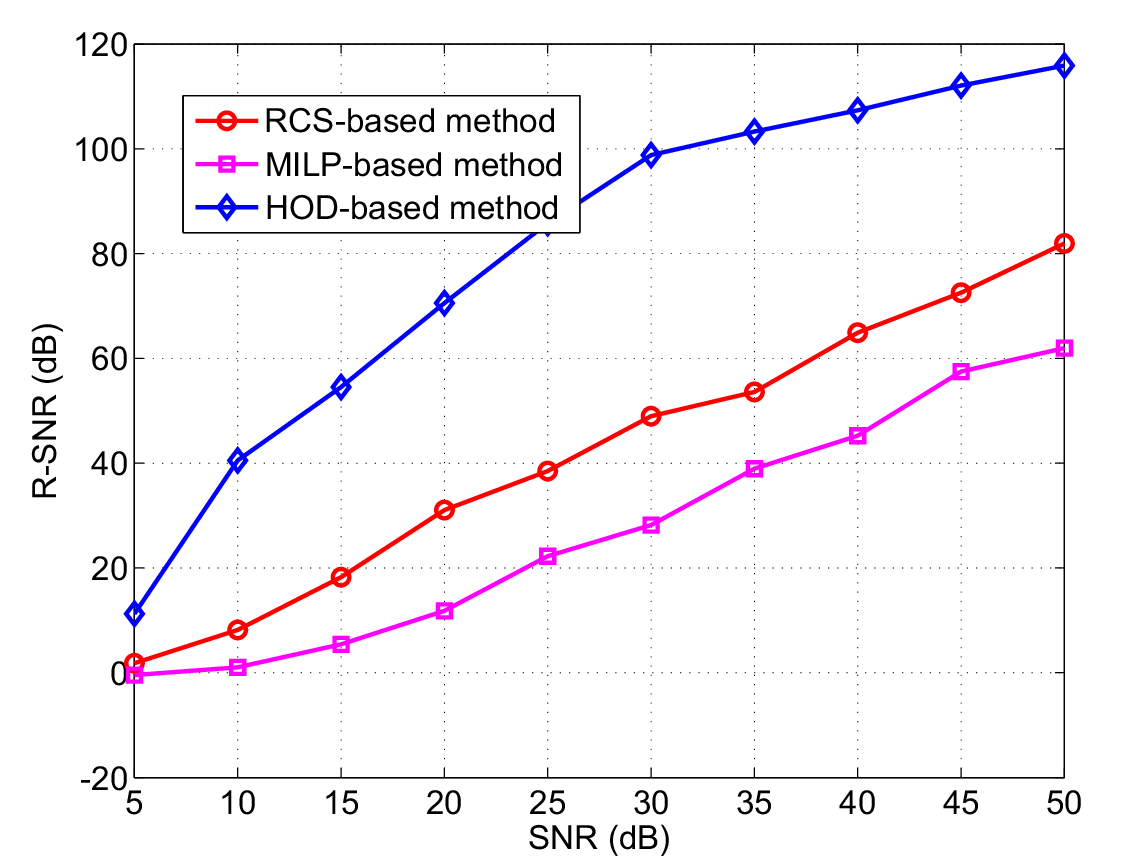}}}\
{\subfloat[$\Delta T = 0.014$s]{\includegraphics[width=0.32\textwidth]{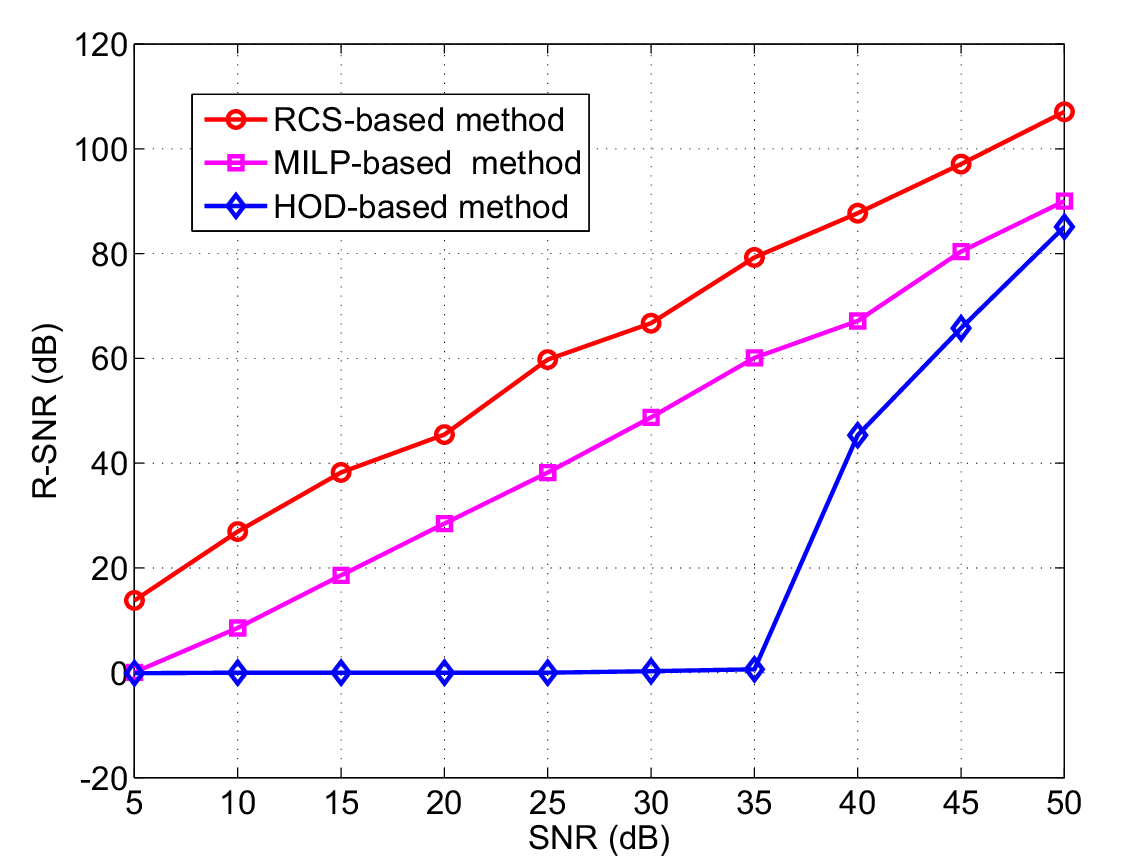}}}\
{\subfloat[$\Delta T = 0.024$s]{\includegraphics[width=0.32\textwidth]{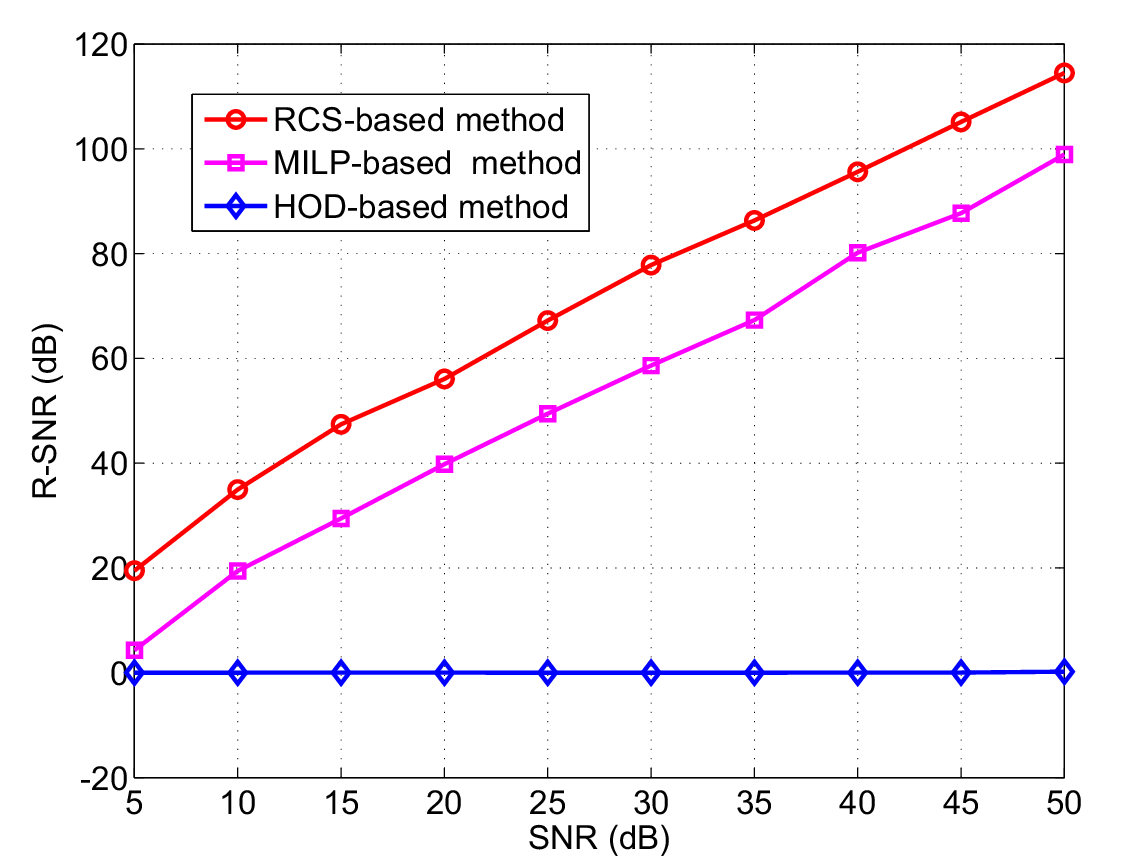}}}
\caption{The NMSE and the R-SNR along with SNR when $M = 400$. } \label{f2}
\end{figure*}

\subsection{Theoretic Analysis}
In this subsection, we examine the condition under which the
integer $\tilde {e}_m=(\tilde {x}_m-\tilde {f}_m)/2\lambda$ is
confined to be in the set $\{0,\pm 1 \}$. We have the following
theorem with respect to the sinusoidal mixture signal defined in
(\ref{eqn1}).

\begin{theorem}
\label{t2} Consider the ensemble of signals $f(t)$ that have a
form of (\ref{eqn1}). If the sampling interval satisfies
\begin{align}
\Delta T < \frac{1}{\omega}\left( \frac{\bar
\beta}{2\lambda}\right)^{-1} \label{cond_t20}
\end{align}
then we have $\tilde f_m - \tilde x_m = 2\lambda \tilde e_m $, where
$\tilde e_m$ is an integer which belongs to the set $\{0,\pm 1\}$.
\end{theorem}
\begin{IEEEproof}
See Appendix~\ref{appC}.
\end{IEEEproof}

Theorem~\ref{t2} indicates that, when the sampling rate is higher
than the threshold in~\eqref{cond_t20}, we can guarantee that
$\tilde {e}_m=(\tilde{x}_m-\tilde{f}_m)/2\lambda$ is confined to be in
the set $\{0,\pm 1 \}$. Similarly, we can rewrite the required
sampling interval as
\begin{align}
\Delta T < \frac{1}{\omega}\left( \frac{\bar
\beta}{2\lambda}\right)^{-1}\triangleq \frac{1}{2\omega \eta_2}
\label{eqn3}
\end{align}
where $\eta_2$ is given by
\begin{align}
\eta_2 \triangleq \frac{\bar N}{2}
\end{align}
in which $\bar N$ is the effective folding number. It is clear
that $\eta_2$ is smaller than $\eta_1$ since $\tau$,
defined before (\ref{dt_p}), should be greater than $0.5$ in
general. Such a relationship between $\eta_1$ and
$\eta_2$ means that the sampling rate condition~\eqref{eqn3} is
less restrictive than condition~\eqref{dt_p}.

\section{Simulation Results}
In this section, we provide simulation results to illustrate the
performance of the proposed first-order difference-based methods,
namely, the RCS-based method and the MILP-based
method. The HOD-based method discussed in Section IV is also
included as a benchmark to show the effectiveness of the proposed
methods. For the RCS-based method, the sparse
Bayesian learning method~\cite{WipfRao04} is employed to solve the
formulated robust compressed sensing problem.

\subsection{Signal Recovery Performance}
We consider $K=3$ sinusoidal components with frequencies are $\{0.4\pi,1.0\pi,1.8\pi\}$, and the number of discretized grid points, $P$, is set to $20$. For each
realization, the complex amplitudes of these
three sinusoidal components are set to ensure $B = 13.6$ and $\bar \beta =
4$. The sampling interval required by each method can be
accordingly determined:
\begin{align}
\Delta T_{\text{HOD}} &\le \frac{1}{2\omega e} = 0.0325 \label{cond_hoder}\\
\Delta T_{\text{RCS}} &\le \frac{1}{2\omega \eta_1} = 0.0884\lambda(1-\tau) \label{cond_rspl}\\
\Delta T_{\text{MILP}} &\le \frac{1}{2\omega \eta_2} =
0.0884\lambda \label{cond_milp}
\end{align}
For the HOD based method, the order of difference has to satisfy the
following condition
\begin{align}
N\ge \left\lceil \frac{\log(0.0735\lambda)}{\log(15.3715\Delta
T)} \right\rceil \label{order}
\end{align}
where $\Delta T \le \Delta T_{\text{HOD}}$ denotes the sampling interval for the HOD based method.
%we uniformly discrete $(0,2\pi]$ into a set of $20$ grid points.

In our simulations, the initial phase of each sinusoidal component
is randomly selected from $(0,2\pi]$. The signal-to-noise ratio
(SNR) is defined as $\text{SNR} = \mathbb{E}(\|\bm
z\|^2)/\mathbb{E}(\|\bm v\|^2)$. The range parameter is set to
$\lambda=1$ throughout our simulations. The
normalized mean square error (NMSE) of $\bm \alpha$ and the
reconstruction SNR (R-SNR) are used as two metrics to evaluate the
estimation accuracy of the complex amplitudes and the frequencies.
These two metrics are defined respectively as
\begin{align}
\text{NMSE}&\triangleq  \frac{\mathbb{E}(\|\bm \alpha-\bm
{\hat\alpha}\|^2)}{\mathbb{E}(\|\bm \alpha\|^2)} \label{nmse}\\
\text{R-SNR}& \triangleq 20\log_{10}\left(\frac{\|\bm y\|^2}{\|\bm y - \hat{\bm y}\|^2}\right)
\end{align}
where $\{\boldsymbol{\alpha},\bm y\}$ and
$\{\boldsymbol{\hat{\alpha}},\bm{\hat y}\}$ denote the true
parameters and the estimated ones, respectively.

\begin{comment}
\begin{figure*}[!t]
\centering
\subfloat {\includegraphics[width=0.32\textwidth]{f41.eps}}\
\subfloat {\includegraphics[width=0.32\textwidth]{f42.eps}}
\subfloat {\includegraphics[width=0.32\textwidth]{f43.eps}}
\caption{The NMSE, the R-SNR as well as the success rate along
with $\Delta T$ when $M=400$ and $\text{SNR}=30$dB.} \label{f3}
\end{figure*}
\end{comment}

The performance of respective methods along with
{SNR} is plotted in Fig.~\ref{f2}, where three different sampling
intervals are employed. We see that the HOD-based method achieves
the best performance when $\Delta T=0.004$s. This is because for
the HOD-based method, the required order of difference is $N=1$
when $\Delta T=0.004$s (cf.~\eqref{order}), in which case the
HOD-based method reduces to a first-order difference method.
Nevertheless, when the sampling interval increases to $\Delta
T=0.014$s and $\Delta T=0.024$s, the required orders of difference
for the HOD-based method are $2$ and $3$, respectively, in which
case the HOD-based method incurs a significant performance
degradation. Specifically, when $\Delta T=0.014$s, the HOD-based
method only works for extremely large SNRs (i.e.,
$\text{SNR}\ge40$dB). In contrast, the proposed first-order
difference-based methods achieve satisfactory performance across
different sampling intervals, ensuring a reliable
estimation even under low SNR conditions.

\begin{figure}[!t]
\centering
\subfloat {\includegraphics[width=0.46\textwidth]{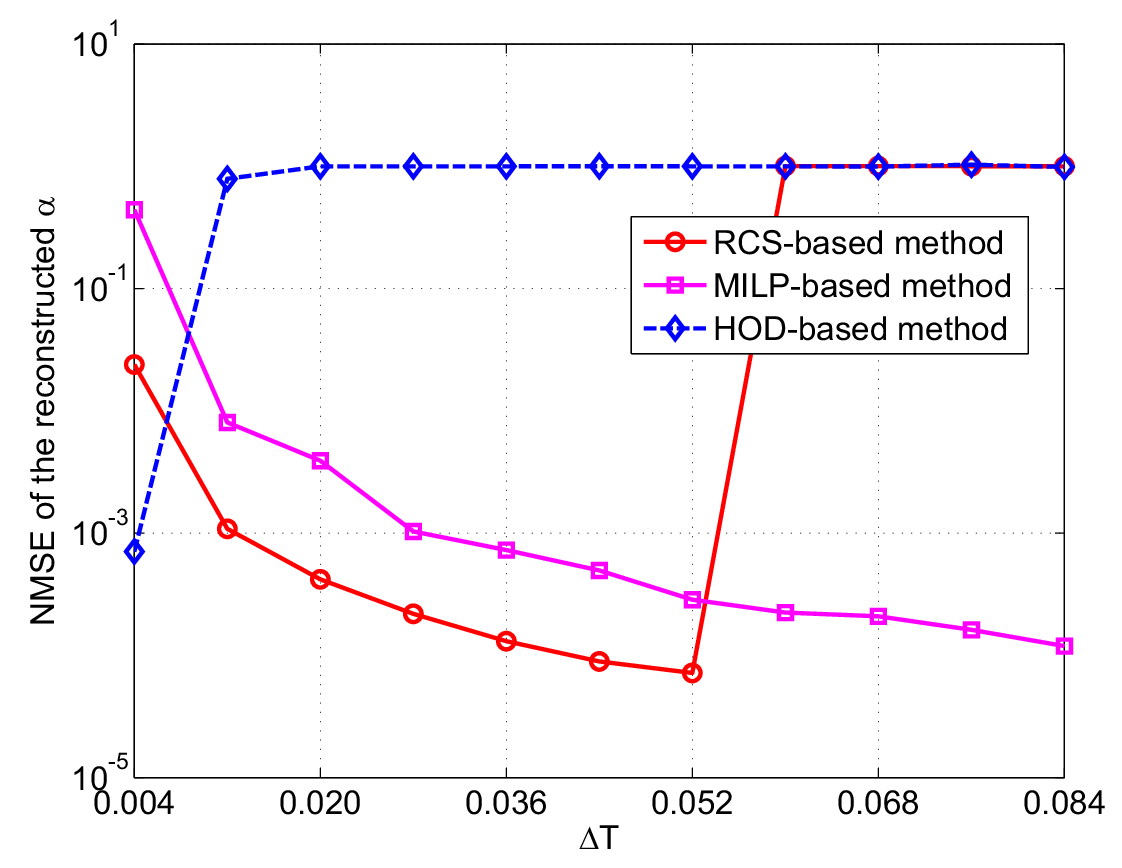}} \
\subfloat {\includegraphics[width=0.46\textwidth]{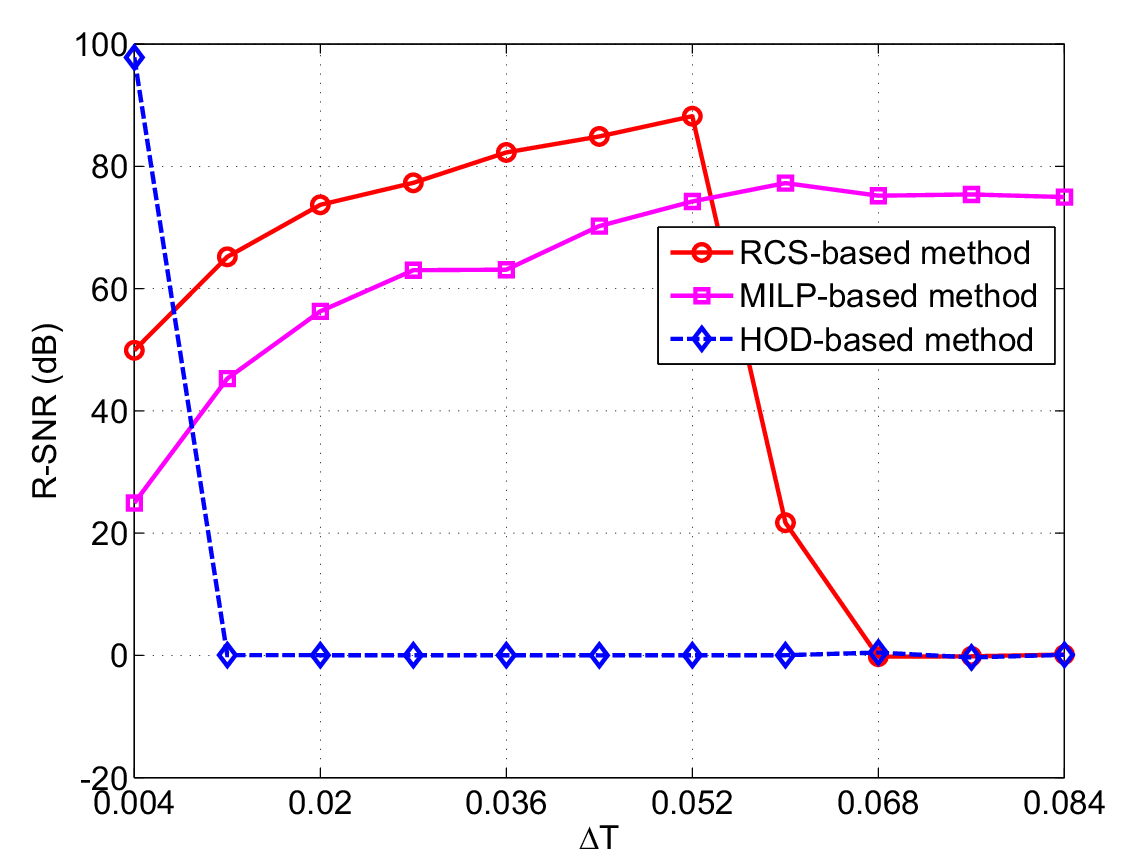}}
\caption{The NMSE and the R-SNR along
with $\Delta T$ when $M=400$ and $\text{SNR}=30$dB.} \label{f3}
\end{figure}

Fig.~\ref{f3} depicts the NMSEs and the R-SNR of respective
methods as the sampling interval $\Delta T$ varies from $0.004$s
to $0.084$s with a stepsize of $0.008$s, where $\text{SNR}$ is set to $30$dB. It can be seen
that the HOD-based method works well only when the sampling
interval is as low as $0.004$s. This is because the required order
of difference is $N=1$ in this case. When the sampling interval
increases to $0.02$s, although it still meets the
condition~\eqref{cond_hoder}, the order of difference required by
the HOD-based method is greater than $1$ according
to~\eqref{order}. Hence the HOD-based method suffers severe
performance degradation due to the shrinkage effect on the signal.
In contrast, for the proposed first-order difference-based
methods, we see that they are able to achieve superior recovery
performance even when a smaller sampling rate is employed. In
particular, the MILP-based method can provide reliable performance
across different values of $\Delta T$ as long as the condition
in~\eqref{cond_milp} is met. The RCS-based method yields a better
estimation accuracy than the MILP-based method when $\Delta T\leq
0.052\text{s}$. When $\Delta T$ further increases,
the performance of the RCS-based method degrades. Specifically,
when $\Delta T = 0.052$s, it can be calculated that the
probability of $\text{Pr}(\tilde f_m = \tilde x_m)$ is no less
than $41.2\%$ according to~\eqref{ppr}. Due to relaxation used in
our derivations, this probability is pessimistic. In
fact, our experiments suggest that when $\Delta T = 0.052$s, the
probability of $\text{Pr}(\tilde f_m = \tilde x_m)$ can reach up to
$90\%$, which implies a majority of elements in $\boldsymbol{s}$
are zeros. Hence the sparse outlier assumption required by the RCS-based
method is well satisfied, and in this case the RCS-based
method outperforms the MILP-based method. When $\Delta T $
increases to $0.060$s, our experiments suggest that the
probability of $\text{Pr}(\tilde f_m = \tilde x_m)$ drops below
$80\%$. The sparse outlier assumption is barely met in this case.
Therefore, when $\Delta T \ge 0.060$s, the RCS-based method experiences
a degraded performance compared to the
MILP-based method.

Another interesting observation of Fig.~\ref{f3} is that,
within the region $0.004\text{s}\le \Delta T \le 0.052\text{s}$,
the performance of the RCS-based and MILP-based methods improves
as the sampling interval increases. The is because, when the
number of samples is fixed, a larger sampling interval usually
results in a less coherent sensing matrix with a more favorable
restricted isometry condition, which in turn leads to a better
recovery performance. This can be verified by checking the mutual coherence of the sensing matrix $\bm{\tilde A}$, which tends to decrease as the sampling interval increases.

\begin{figure}[!t]
\centering
\subfloat[$\text{SNR}=20$dB] {\includegraphics[width=0.46\textwidth]{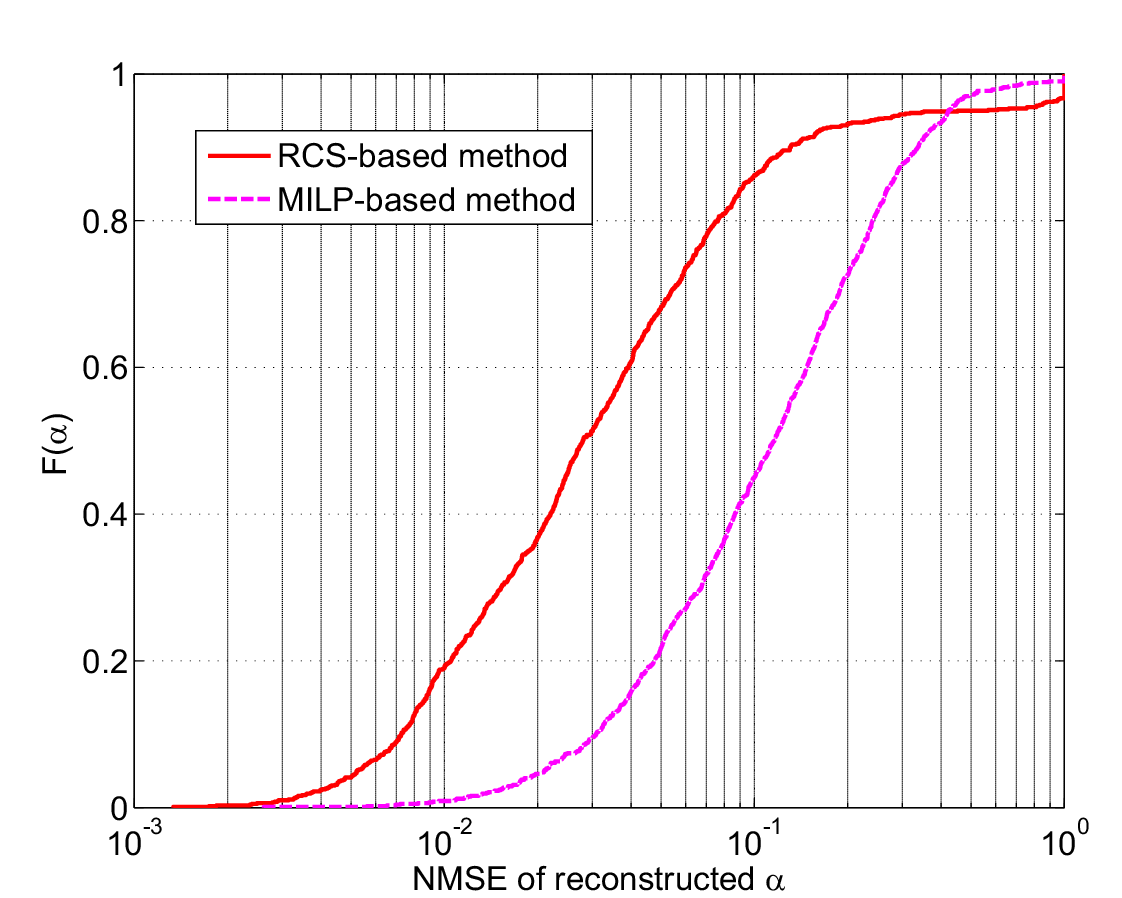}} \
\subfloat[$\text{SNR}=50$dB] {\includegraphics[width=0.46\textwidth]{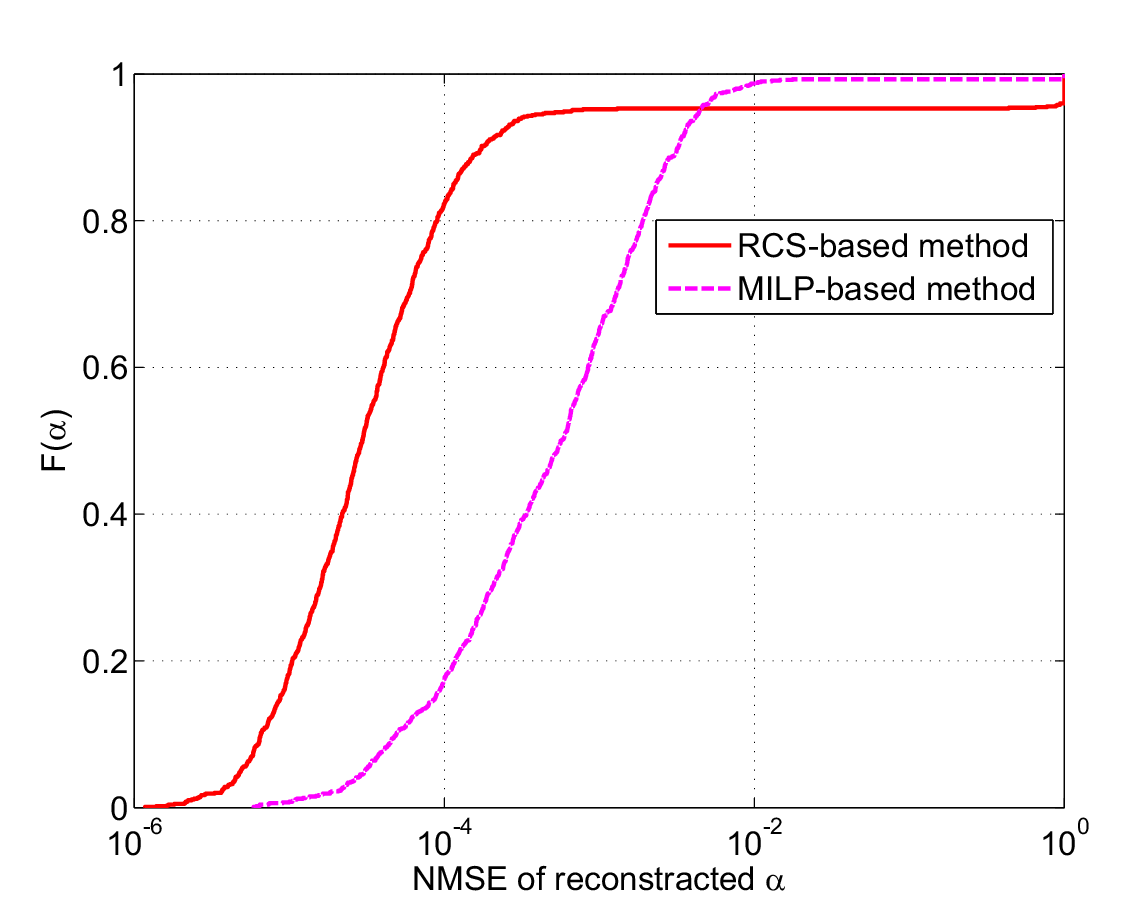}}
\caption{The empirical PDF of the NMSE with $M = 400$ when $\text{SNR}=20$dB and $\text{SNR}=50$dB.} \label{f333}
\end{figure}

In addition, we consider a scenario where the number of sinusoidal components increases to $K=10$. For each independent Monte Carlo run, the frequencies of these sinusoidal components are selected from the pre-defined grids, while their amplitudes are randomly selected from a Gaussian distribution $\mathcal{N}(0,1)$. The sampling interval is set to $\Delta T = 0.027$s and the dynamic range of the modulo ADC is set to $\lambda = 0.5$. We consider two cases, i.e., $\text{SNR} = 20$dB and $\text{SNR} = 50$dB. Results are averaged over $1000$ independent realizations. Since the sampling interval is fixed while both the frequencies and the amplitudes are randomly selected, the sampling conditions required for the RCS-based method or/and the MILP-based method may not be satisfied. In Fig.~\ref{f333}, we illustrate the empirical PDF of the NMSE of these two cases, and we can see that the MILP-based method has a higher chance to succeed compared with the RCS-based method. This is because the MILP-based method has a more relaxed sampling condition compared with the RCS-based method, as shown in~\eqref{eqn3} and~\eqref{dt_p}. This result is consistent with the previous conclusion as seen in Fig.~\ref{f2} and Fig.~\ref{f3}.

\begin{figure}[!t]
\centering
{\includegraphics[width=0.6\textwidth]{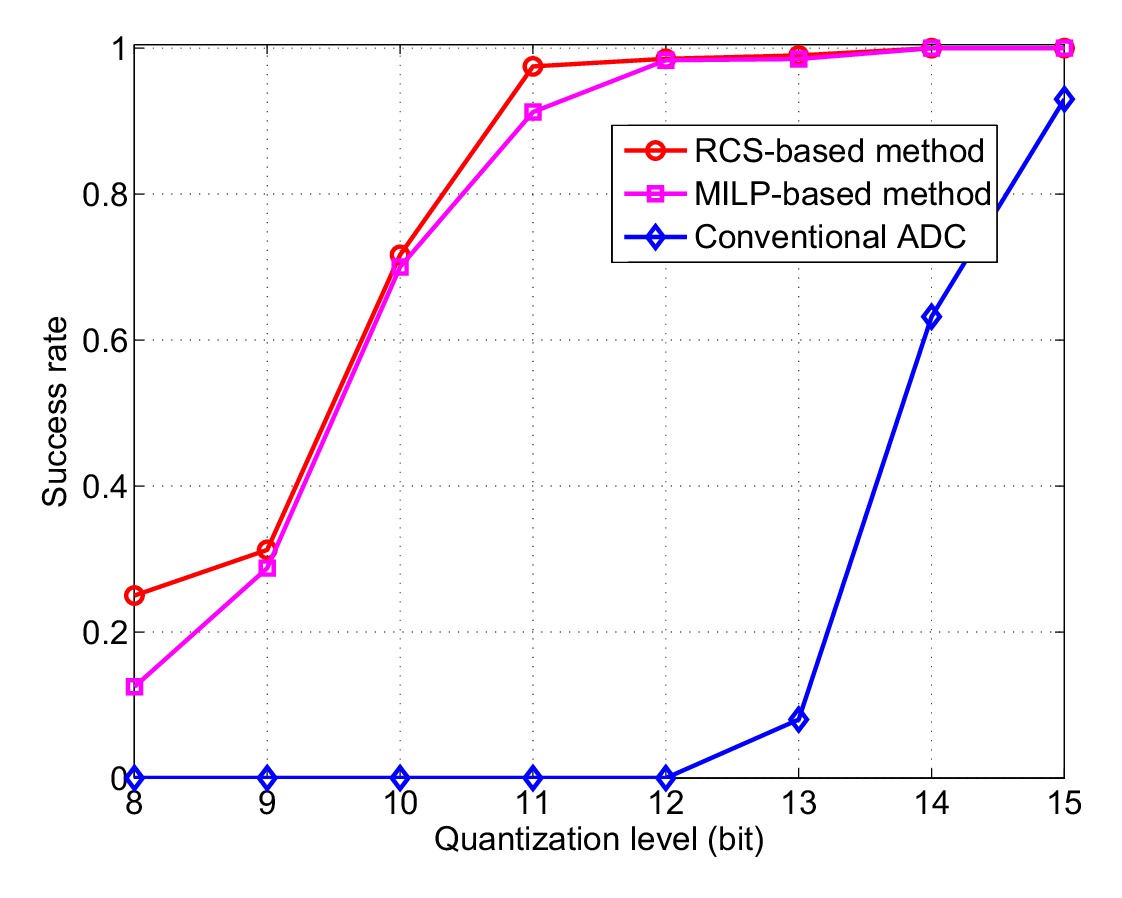}}
\caption{The successful rate with varied quantization levels when $M = 400$.}
\label{f4}
\end{figure}

\begin{figure*}[!t]
\centering
\subfloat[]{\includegraphics[width=0.32\textwidth]{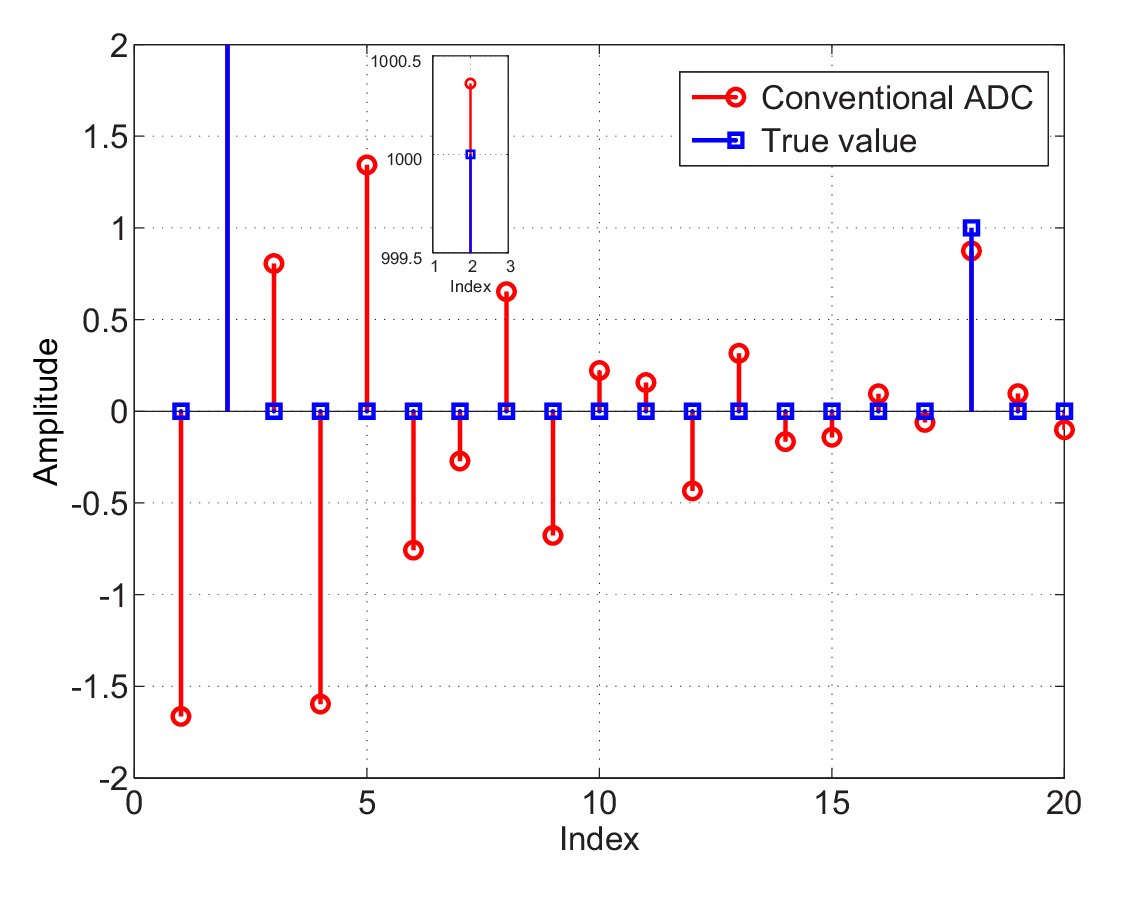}}\
\subfloat[]{\includegraphics[width=0.32\textwidth]{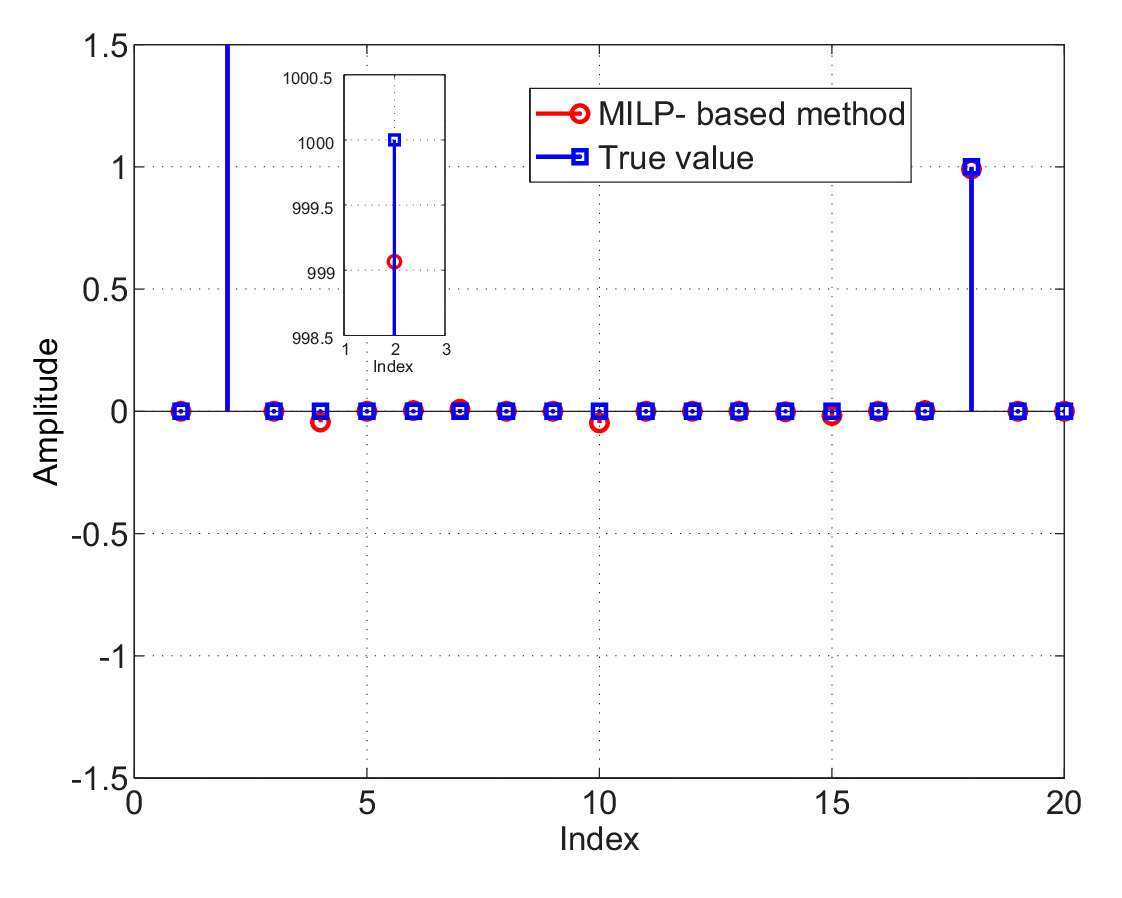}}\
\subfloat[]{\includegraphics[width=0.32\textwidth]{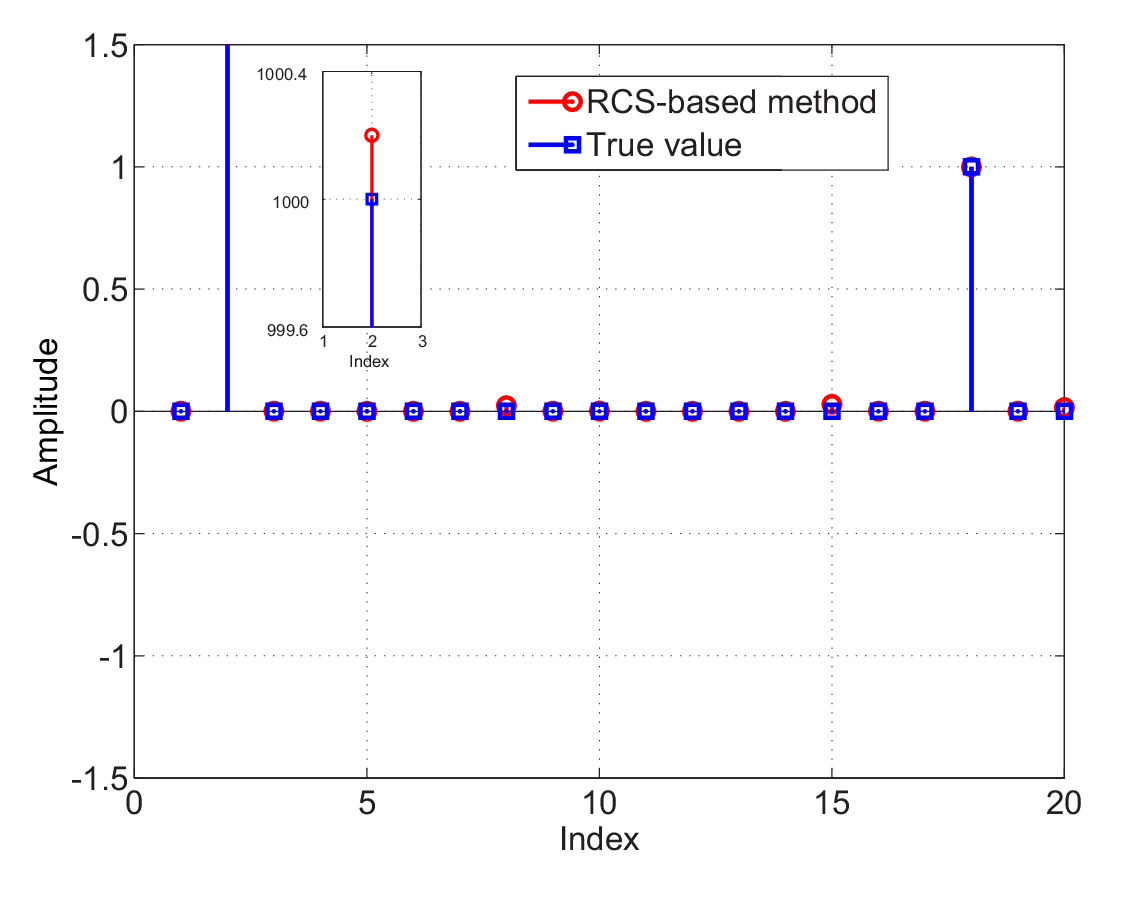}}
\caption{The estimated $\bm \alpha$ of different methods in one
specific run when $M = 600$ and the quantization level is $ 10$bit.} \label{f7}
\end{figure*}

\subsection{Results with High Dynamic Ranges}
The unlimited sensing framework is a potential solution to deal
with a difficulty in wireless communications caused by the near-far effect, where due to
the limited dynamic range, a weak signal cannot be detected in
the presence of a strong background signal. In this experiment, we
illustrate the performance of the proposed methods in scenarios
with a high dynamic range. To this end, we consider $K=2$
sinusoidal components, and the corresponding parameters are given
as
\begin{align*}
&|\alpha_1| = 1000,\ |\alpha_2| = 1, \ \omega_1 = 0.2\pi,\ \omega_2 = 1.8\pi
\end{align*}
The ratio of the power of the strong signal to the
power of the weak signal is $60$dB. To high-fidelity data-acquisition without saturation, a conventional ADC would
need a large number of quantization bits. Otherwise the weak
signal will be buried beneath the quantization noise level and
cannot be effectively detected. Modulo ADCs do not suffer from
this issue. Theoretically, modulo ADCs can achieve an
unlimited dynamic range by folding the signal back to its range
interval. Here we also consider the quantization noise for
modulo ADCs. Unlike conventional ADCs, modulo ADCs
quantize its modulo samples instead of the original input samples.
In our simulations, we set $\lambda=10$ and $\Delta T=0.01$s.

To evaluate the performance of detecting the weak signal amidst the
strong background signal, we use the success rate as a metric: a trial is considered successful if both of the following conditions are met:
\begin{enumerate}
\item The support set of the sparse signal $\bm \alpha$ is correctly found.
In general, most components of the estimated $\bm \alpha$ will not
exactly equal $0$ due to the quantization noise. We set those
components to zero if their amplitude is smaller than $0.1$.
\item The estimation error of each non-zero element,
defined as $|\bm {\hat \alpha}(i)-\bm \alpha(i)|/|\bm \alpha(i)|$, is no greater than $0.15$.
\end{enumerate}

Fig.~\ref{f4} plots the success rates of respective methods as a
function of the number of quantization bits. We see that the
success rates of both proposed methods increase when more
quantization bits are used. Specifically, a success rate of $1$
can be achieved when $11$ bits are employed to quantize the modulo
samples. In contrast, the conventional ADC needs $15$ quantization
bits to achieve a decent success rate.
In Fig.~\ref{f7}, we plot the estimated $\bm \alpha$ for a
particular realization when $M = 600$ and the number of
quantization bits is set to $10$. It can be seen that while the proposed modulo ADC based methods have no difficulty in recovering both the strong and weak signals, the conventional ADC cannot distinguish the weak signal from many false signals it produces.

%We compare the conventional ADC and self-reset ADC under different
%quantization levels. The dynamic range of the conventional ADC is
%$[-1024,1024]$, while that of the self-reset ADC is $[-10,10]$
%(i.e., $\lambda =10$). In the simulation, we set $\Delta t =
%0.01$s. We implement $200$ Monte Carlo runs to calculate the
%simulation results.

\section{Conclusions}
In this work, we studied the problem of LSE in an unlimited sensing framework, which employs modulo ADCs
to nonlinearly map the received signal to modulo samples to avoid
clipping or saturation problems. We first introduced a HOD based
approach, which is sensitive to noise and lacks satisfactory performance in the presence of noise. To
overcome this difficulty, we investigated the properties of
the first-order difference of modulo samples, and developed two
first-order difference-based methods, namely, the RCS-based and
the MILP-based methods. Simulation results show that both methods
are robust against noise and achieve a significant performance
improvement over the HOD-based method. In practice, due to implementation imperfections of
modulo ADCs such as hysteresis effects~\cite{FlorescuKrahmer22}
and impulsive noises~\cite{Bhandari22}, the folding numbers might
not be integers. How to extend our methods to deal with these
nonlinearities is an important topic for future
investigation.

%\useRomanappendicesfalse
\appendices

\section{Discussion On~\eqref{model_pm}}
\label{dis} According to~\cite{WidrowKollar96}, for
an input signal $x$ to be quantized, if the characteristic
function (CF) of the random variable $x$, denoted as
$\Phi_x(u)$, is ``bandlimited'', i.e., it satisfies
\begin{align}
\Phi_x(u) = 0 \quad  \ |u|\ge\frac{2\pi}{2\lambda}
\end{align}
where $2\lambda$ is the quantization step size, then the
quantization noise follows a uniform distribution. This
condition is restrictive, as most frequently encountered signals
such as Gaussian signals do not have perfectly bandlimited CFs.
Therefore, an approximate condition is widely accepted.
Specifically, if $\Phi_x(u)$ is {\emph{approximately
bandlimited}} for some quantization step size, i.e.,
\begin{align}
\Phi_x(u) \approx 0 \qquad \ |u|\ge\frac{2\pi}{2\lambda}
\label{app_bandlimited}
\end{align}
then the quantization noise can be approximately modeled as a
uniform distribution.

\begin{figure}[!t]
\centering
{\includegraphics[width=0.60\textwidth]{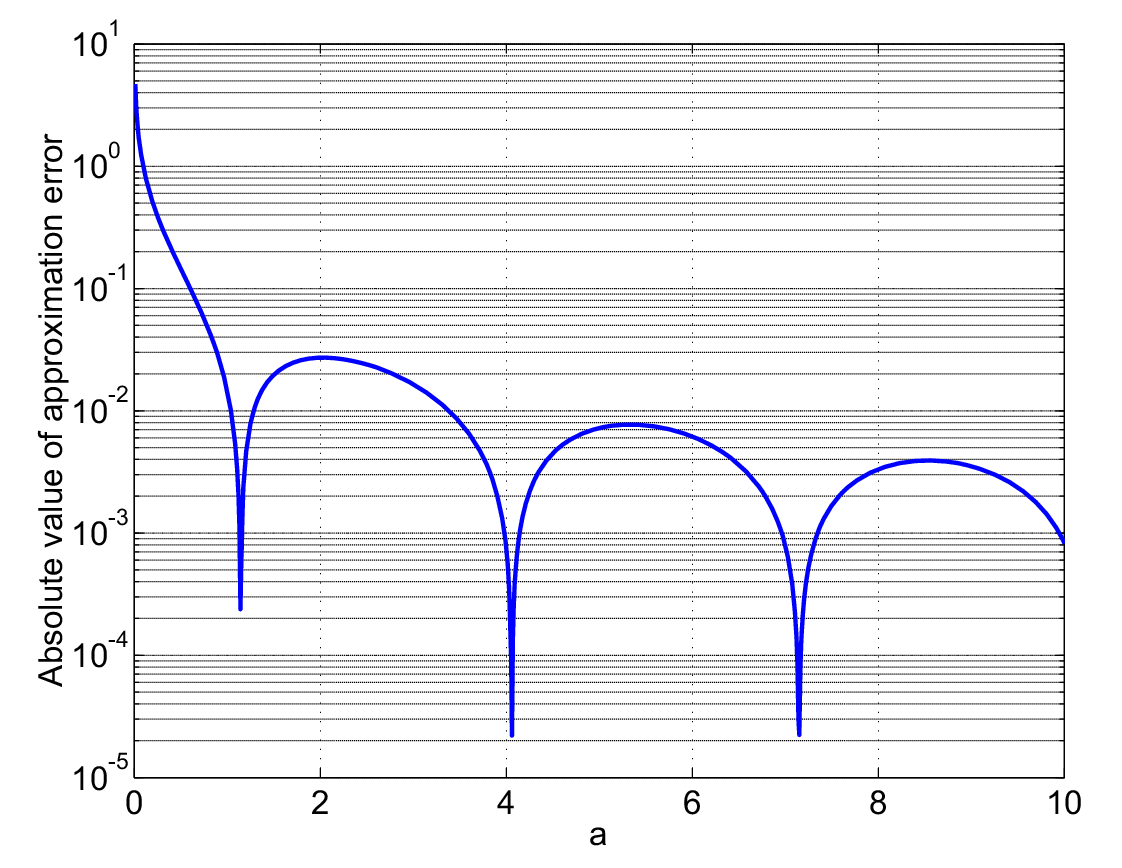}}
\caption{The absolute value of the approximation error of~\eqref{app_bess} as a function of $a$.}
\label{f_app_bess}
\end{figure}

In our problem, we assume that the phase of the $k$th component,
i.e., $f_k \triangleq \beta_k\cos(\omega_kt+\varphi_k)$, is
uniform in the interval $(-\pi,\pi)$. Therefore, the probability
density function of $f_k$ is given by~\cite{Barakat74,BarakatCole79,EditionPapoulis02}
\begin{align}
p_k(f_k) = \left\{
\begin{array}{ll}
\frac{1}{\pi\sqrt{\beta_k^2 - f_k^2}},& |f_k|\le |\beta_k| \\
\ & \ \\
0,& |f_k|> |\beta_k|\\
\end{array}
\right.
\end{align}
Accordingly, the CF of the random variable $f_k$ is given
by~\cite{Barakat74}
\begin{align}
\Phi_k(u) = \mathcal{J}_0\big(|\beta_k|u\big)
\end{align}
where $\mathcal{J}_0(\cdot)$ denotes the Bessel function of the first kind. Note
that the CF of the sum of $K$ independent random variables is the
product of individual CFs associated with these random variables.
Therefore, the CF of $f(t)$ can be calculated as
\begin{align}
\Phi_f(u) = \prod_{k=1}^K \Phi_k(u) = \prod_{k=1}^K \mathcal{J}_0\big(|\beta_k|u\big)
\end{align}
Note that the Bessel function $|\mathcal{J}_0(a)|< 1$ when $a \ne
0$. Hence we have
\begin{align}
\Big|\Phi_f(u)\Big| <  \Big|\mathcal{J}_0\big(|\beta_k|u\big)\Big|,\ \forall k\in\{1,\cdots,K\}
\label{scale}
\end{align}
where $u \ne 0$. In addition, the Bessel function
$\mathcal{J}_0(a)$ is an even function and looks like a decaying sinusoid that decays proportionally to
$1/\sqrt{|a|}$. Specifically, for $|a| \gg 1/4$,
$\mathcal{J}_0(a)$ can be approximated
as~\cite{AbramowitzStegun64}
\begin{align}
\mathcal{J}_0(a) \approx \sqrt{\frac{2}{\pi |a| }} \cos(|a| -\frac{\pi}{4})
\label{app_bess}
\end{align}
The absolute value of the approximation error in~\eqref{app_bess} is shown in Fig.~\ref{f_app_bess}, where we can see that, when $a\ge 2$, the absolute error is less than $0.03$. In the following, we suppose that~\eqref{app_bess} holds exactly (i.e., the approximation error is negligible) when $|a|\ge 2$.

Setting $\tilde \beta \triangleq \max\{|\beta_{k}|\}$,
$\mathcal{J}_0(\tilde \beta u)$ has a faster decaying rate than $\mathcal{J}_0(|\beta_k u|)$. When $|\tilde \beta u |\ge 2$, we have
\begin{align}
\big|\Phi_f(u)\big| &\overset{(a)}{<} \Big|\mathcal{J}_0(\tilde \beta u)\Big|\notag\\
& \overset{(b)}{\approx}\left|\sqrt{\frac{2}{\pi |\tilde \beta
u|}}\cos(|\tilde \beta u| -\frac{\pi}{4})\right| \le \sqrt{\frac{2}{\pi \tilde \beta |u|}}
\label{app_cf}
\end{align}
where $(a)$ is directly from~\eqref{scale}, and in $(b)$ we employ the approximation in~\eqref{app_bess}. From~\eqref{app_cf} we
know that if
\begin{align}
|u| \ge \max\left\{\frac{2}{\pi \tilde \beta \varsigma^2},\frac{2}{\tilde \beta}\right\}
\label{cond_u}
\end{align}
then the following inequality holds
\begin{align}
\big|\Phi_f(u)\big|\le \varsigma
\end{align}
where $\varsigma$ is a small positive value. Generally, $\varsigma$ is set sufficiently small such that $\pi \varsigma^2 \le 1$ (specifically, $\varsigma $ is required to be no greater than $0.564$). Therefore,  the condition in~\eqref{cond_u} is equivalent to
\begin{align}
|u| \ge \frac{2}{\pi \tilde \beta \varsigma^2}
\end{align}
Based on the above result, to ensure
that the approximately bandlimited condition~\eqref{app_bandlimited} holds, the following condition should be
satisfied
\begin{align}
|u| \ge \frac{2\pi}{2\lambda} \ge \frac{2}{\pi \tilde \beta \varsigma^2}
\label{cond_u_final}
\end{align}
The above condition implies
\begin{align}
2\lambda \le (\pi\varsigma)^2 \tilde \beta
\label{cond_lambda}
\end{align}
Considering the fact that $B \triangleq \sum_{k=1}^K|\beta_k|$ and $\tilde \beta \triangleq \max\{|\beta_{k}|\}$, we have
\begin{align}
2\lambda \le (\pi\varsigma)^2 \frac{B}{K}
\label{cond_lambda_final}
\end{align}
to ensure~\eqref{cond_u_final}.

From~\eqref{cond_lambda_final}, we know that the operation range of the
modulo ADC, i.e., $2\lambda$, should be smaller than
$(\pi\varsigma)^2 B/K$, such that $x_m$ can be approximately modeled
as a uniform distribution. This is equivalent to the condition
that the maximum folding number, defined by $\mathcal{K}\triangleq
{B}/{(2\lambda)}$, should satisfy the following inequality:
\begin{align}
\mathcal{K} \ge \frac{K}{(\pi\varsigma)^2}
\label{cond_bandlimited}
\end{align}
If we set $K=4$ and $\varsigma = 0.1$, $\mathcal{K}$ should be no less than $40$. This condition is
usually met considering the fact that the unlimited sensing
framework aims to deal with high dynamic range problems.

%The reason is twofold. For one thing, such a condition is not a
%restrictive condition since we have several relaxation operations
%during obtaining such a condition, i.e.,~\eqref{app_cf}
%and~\eqref{cond_lambda}. For another, even
%though~\eqref{cond_bandlimited} is conservative,

\section{Proof of Theorem~\ref{t1}}
\label{appA} By the modulo decomposition property~\cite{BhandariKrahmer20}, we
have
\begin{align}
x_{m+1} = f_{m+1} + 2\lambda e_{m+1}, \
x_{m} = f_{m} + 2\lambda e_{m}\notag
\end{align}
where $e_{m+1},e_{m}\in\mathbb{Z}$. Therefore, we have
\begin{align}
\tilde x_m & =\tilde f_m - 2\lambda (e_{m} - e_{m+1})\notag\\
&\triangleq \tilde f_m - 2\lambda \tilde e_m \label{fod_modu_comp}
\end{align}
where $\tilde e_m\triangleq (e_{m} - e_{m+1}) \in\mathbb{Z}$. Therefore,
$\tilde x_m$ can be uniquely represented by a sum of $\tilde f_m$ and
an unknown constant. In scenarios where this constant equals $0$, $\tilde x_m$ is equivalent to $\tilde f_m$. Therefore, we have
\begin{align}
\text{Pr}(\tilde f_m = \tilde x_m) \Leftrightarrow \text{Pr}(\tilde e_m
= 0)
\end{align}
By the definition of the modulo operation in~\eqref{folding_oper},
we have
\begin{align}
x_{m+1} = f_{m+1} - 2\lambda \left\lfloor \frac{f_{m+1} }{2\lambda} +
\frac{1}{2}\right\rfloor
\end{align}
which indicates that
\begin{align}
e_{m+1} = - \left\lfloor \frac{f_{m+1} }{2\lambda} +
\frac{1}{2}\right\rfloor \label{et}
\end{align}
Similarly, $e_{m}$ can be expressed as
\begin{align}
e_{m} = - \left\lfloor \frac{f_{m} }{2\lambda} +
\frac{1}{2}\right\rfloor \label{etm1}
\end{align}
Combining~\eqref{et} and~\eqref{etm1} results in
\begin{align}
\tilde e_m &\triangleq e_{m}-e_{m+1}\notag\\
&=\left(\left\lfloor \frac{f_{m+1}}{2\lambda} + \frac{1}{2}\right\rfloor
- \left\lfloor \frac{f_{m}}{2\lambda} + \frac{1}{2}\right\rfloor\right)\notag\\
& = \Bigg(\left\lfloor \frac{f_{m} }{2\lambda} + \frac{1}{2} +
\frac{f_{m+1} - f_{m}}{2\lambda}\right\rfloor  - \left\lfloor
\frac{f_{m}}{2\lambda} + \frac{1}{2}\right\rfloor\Bigg)
\label{et_proof}
\end{align}
Since $f(t)$ is continuous and differentiable on the interval
$[(t-1)\Delta T,t\Delta T]$, based on the mean value theorem,
there exists a value $\zeta\in[(t-1)\Delta T,t\Delta T]$ such that
\begin{align}
f'(\zeta) = \frac{f_{m+1} - f_{m}}{\Delta T} \label{et_pf2}
\end{align}
where $f'(t)$ is the derivative of $f(t)$.
Substituting~\eqref{et_pf2} into~\eqref{et_proof}, we arrive at
\begin{align}
\tilde e_m = \left\lfloor b \right\rfloor - \left\lfloor a
\right\rfloor \label{baret}
\end{align}
where $a$ and $b$ are, respectively, defined as
\begin{align*}
a&\triangleq \frac{f_{m}}{2\lambda} + \frac{1}{2}\\
b&\triangleq \frac{f_{m} }{2\lambda} + \frac{1}{2} + \frac{{f}'(\zeta)\Delta T}{2\lambda}
\end{align*}
By these definitions we have
\begin{align}
b & = a + \frac{{f}'(\zeta)\Delta T}{2\lambda} \triangleq a + c
\end{align}
where $c$ is defined as
\begin{align}
c \triangleq \frac{{f}'(\zeta)\Delta T}{2\lambda}
\end{align}
The derivative of $f(t)$ is given by
\begin{align}
f'(t) = -\sum_{k=1}^K\beta_k\omega_k\sin(\omega_k t+\varphi_k)
\end{align}
which leads to the following bound
\begin{align}
|c| \le \frac{\Delta T\sum_{k=1}^K|\beta_k\omega_k|}{2\lambda} =
\frac{\Delta T\omega\bar \beta}{2\lambda}\triangleq \delta
\label{p2_2}
\end{align}

As in~\eqref{exp_fm}, $f_m$ can be expressed as
\begin{align}
f_m &= 2\lambda \left\lfloor
\frac{f_m}{2\lambda}+\frac{1}{2}\right\rfloor + x_m = 2\lambda \left\lfloor a \right\rfloor + x_m
\label{quan0}
\end{align}
where $x_m$ can be modeled as a uniform distribution:
\begin{align}
x_m\sim\mathbb{U}(-\lambda,\lambda)
\end{align}
In addition, from the quantization theory in~\cite{WidrowKollar96}, $x_m$ is statically independent of $f_m$. Taking a simple algebraic operation
on~\eqref{quan0} results in
\begin{align}
\frac{f_m}{2\lambda} +\frac{1}{2} = \left\lfloor
a\right\rfloor + \tilde{q}_m
\label{quan2}
\end{align}
where $\tilde{q}_m\triangleq x_m/(2\lambda) + 1/2$ also follows a
uniform distribution, i.e., $\tilde {q}_m\sim\mathbb{U}(0,1)$.
Recalling the definition of $a$, we have
\begin{align}
a = \left\lfloor a \right\rfloor + \tilde q_m
\end{align}
Therefore, $a$ can be assumed to follow a uniform distribution
over the interval $(\left\lfloor a\right\rfloor,\left\lfloor
a\right\rfloor+1)$:
\begin{align}
p_A(a)\sim\mathbb{U}\Big(\left\lfloor
a\right\rfloor,\left\lfloor
a\right\rfloor +1\Big) \label{pa}
\end{align}

For $c$, it can be modeled as a
random variable and its distribution $p_C(c)$ is given as (as we
have $|c|\leq\delta$)
\begin{align}
p_C(c) = \left\{
\begin{array}{ll}
0,&c<-\delta \  \text{or}\ c> \delta\\
p_C(c),&-\delta \le c \le \delta
\end{array}
\right.
\end{align}
where $p_C(c)$ is an arbitrary distribution that satisfies
\begin{align}
\int_{-\delta}^{\delta} p_C(c) dc = 1
\end{align}
Furthermore, it is reasonable to assume that $a$
and $c$ are independent of each other. This is because the
quantization noise, $\tilde{q}_m$, is independent of $f(t)$, and
$c$ is only related to the derivative of $f(t)$. In this case,
the probability density function (PDF) of $b$ is given by the
convolution of $p_A(a)$ and $p_C(c)$, i.e.,
\begin{align}
p_B(b) = \int_{-\infty}^{\infty}{p_C(b-a)p_A(a)}da
\label{pb}
\end{align}

Recall that the sampling interval $\Delta T$ satisfies
\begin{align}
\Delta T \le \frac{1}{2\omega}\left(\frac{\bar
\beta}{2\lambda}\right)^{-1}
\end{align}
Hence we have
\begin{align}
\delta = \frac{\Delta T\omega\bar \beta}{2\lambda}\le
\frac{1}{2}
\end{align}
and consequently, $\left\lfloor
a\right\rfloor +1- \delta \ge \lfloor a\rfloor
+ \delta$. Therefore, $p_B(b)$ can be calculated as
\begin{align}
 p_B(b) =\left\{
\begin{array}{ll}
0,&b < \lfloor a\rfloor - \delta\\
F_C(b - \lfloor a\rfloor),& \lfloor a\rfloor - \delta \le b <\lfloor a\rfloor + \delta\\
1,& \lfloor a\rfloor + \delta \le b < \left\lfloor
a\right\rfloor +1- \delta\\
1-F_C(b-\left\lfloor
a\right\rfloor -1),& \left\lfloor
a\right\rfloor +1- \delta \le b< \left\lfloor
a\right\rfloor +1 + \delta\\
0,& b > \left\lfloor
a\right\rfloor +1 + \delta
\end{array}
\right.
\end{align}
where $F_C(\cdot)$ is the cumulative distribution function (CDF)
of $c$. The derivation of $p_B(b)$ can be found in
Appendix~\ref{appB}. The desired probability is then computed as
\begin{align}
&\text{Pr}(\tilde e_m = 0) \Leftrightarrow \text{Pr}(\lfloor b\rfloor = \lfloor a\rfloor)\notag\\
&= \text{Pr}(\lfloor a\rfloor \le b < \left\lfloor
a\right\rfloor +1)\notag\\
& = \int_{\lfloor a\rfloor}^{\left\lfloor
a\right\rfloor +1}{p_B(b)}db\notag\\
& \triangleq \kappa_1 + \kappa_2 + \kappa_3 \label{ff1}
\end{align}
where $\kappa_1$,~$\kappa_2$ and $\kappa_3$ are respectively
calculated by
\begin{align}
\kappa_1 & = \int_{\lfloor a\rfloor}^{\lfloor a\rfloor +\delta}
{F_c(b-\lfloor a\rfloor)}db= \int_{0}^{\delta} F_C(c) dc \label{k1}\\
\kappa_2 & = \int_{\lfloor a\rfloor+\delta }^{\left\lfloor
a\right\rfloor +1 -\delta}{(1) }db = 1- 2\delta \label{k2} \\
\kappa_3 &= \int_{\left\lfloor
a\right\rfloor +1-\delta}^{\left\lfloor
a\right\rfloor +1}
{\big(1-F_C(b - \left\lfloor
a\right\rfloor -1)\big)}db =\delta - \int_{-\delta}^{0}
F_C(c)dc \label{k3}
\end{align}
Therefore, we have
\begin{align}
\text{Pr}(\tilde e_m = 0) &= \kappa_1 + \kappa_2 + \kappa_3\notag\\
& = 1- \delta + \int_{0}^{\delta} F_C(c) dc -
\int_{-\delta}^{0} F_C(c)dc \label{ff2}
\end{align}
Due to the property of CDF, $F_C(c)$ is non-negative and
non-decreasing over the interval $[-\delta,\delta]$. Therefore,
the following holds:
\begin{align}
\int_{0}^{\delta} F_C(c) dc - \int_{-\delta}^{0} F_C(c)dc
\ge 0 \label{ff3}
\end{align}
Combining~\eqref{ff2} and~\eqref{ff3}, we arrive at our main
result
\begin{align}
\text{Pr}(\tilde f_m=\tilde x_m)\Leftrightarrow\text{Pr}(\tilde e_m = 0
) \ge 1 - \delta = 1- \frac{\Delta T\omega\bar \beta}{2\lambda}
\end{align}
This completes our proof.

\section{Derivation of $p_B(b)$}
\label{appB} The PDF of $b$ can be expressed as
\begin{align}
p_B(b) & = \int_{-\infty}^{\infty}{p_C(b-a)p_A(a)}da\notag\\
& = \int_{\lfloor a\rfloor }^{\left\lfloor
a\right\rfloor +1}{p_C(b-a)}da\notag\\
& \overset{(a)}{=} \int_{b -\left\lfloor
a\right\rfloor -1}^{b -\lfloor
a\rfloor}{p_C(c)}dc \label{pb_app}
\end{align}
where in $(a)$ we utilize the definition $c \triangleq b  - a$. Recall that
$p_C(c)$ is nonzero only within the interval $[-\delta, \delta]$.
Therefore we have $p_B(b) = 0$ if $b - \lfloor a\rfloor <
-\delta$ or $b - \left\lfloor
a\right\rfloor -1 > \delta$ (i.e., $b <
\lfloor a\rfloor -\delta $ or $b >\left\lfloor
a\right\rfloor +1 +\delta $).

On the other hand, when $-\delta \le b  - \left\lfloor
a\right\rfloor - 1 <
\delta$, i.e., $-\delta +  \left\lfloor
a\right\rfloor +1  \le b <\left\lfloor
a\right\rfloor +1 +\delta $, we have
\begin{align}
p_B(b) = \int_{b -\left\lfloor
a\right\rfloor  - 1}^{\delta }{p_C(c)}dc = 1 -
F_C(b -\left\lfloor
a\right\rfloor - 1)
\end{align}
Moreover, when $b -\left\lfloor
a\right\rfloor  - 1 < -\delta$ and $b -
\lfloor a \rfloor \ge \delta$, i.e., $\lfloor a \rfloor +
\delta \le b < \left\lfloor
a\right\rfloor +1  -\delta$, we have
\begin{align}
p_B(b) = \int_{-\delta}^{\delta }{p_C(c)}dc = 1
\end{align}
Also, when $-\delta \le b - \lfloor a \rfloor < \delta$, i.e.,
$\lfloor a \rfloor -\delta \le b < \lfloor a \rfloor +
\delta$, we have
\begin{align}
p_B(b) = \int_{-\delta }^{b -\lfloor a \rfloor}{p_C(c)}dc =
F_C(b -\lfloor a \rfloor)
\end{align}
Summarizing the above results, we arrive at \eqref{pb}.

\section{Proof of Theorem~\ref{t2}}
\label{appC} From~\eqref{baret}, we know that $\tilde e_m$ can be
expressed by
\begin{align}
\tilde e_m = \left\lfloor b \right\rfloor - \left\lfloor a
\right\rfloor
\end{align}
where $b = a + c$. Based on~\eqref{p2_2}, we know that
\begin{align}
|c| \le \delta = \frac{\Delta T \omega \bar \beta}{2\lambda}
\end{align}
Recalling (\ref{cond_t20}), we have
\begin{align}
\|c\|_{\infty} < \frac{ \omega \bar
\beta}{2\lambda}\frac{1}{\omega}\left( \frac{\bar
\beta}{2\lambda}\right)^{-1} = 1
\end{align}
which indicates that $c$ belongs to the set
\begin{align}
-1<c <1 \label{c-t}
\end{align}
On the other hand, we know that
\begin{align}
\lfloor a\rfloor \le a < \lfloor a \rfloor +1 \label{a-t}
\end{align}
Combining~\eqref{c-t} and~\eqref{a-t} we obtain that
\begin{align}
\lfloor a\rfloor - 1< b < \lfloor a \rfloor +2
\end{align}
Therefore $\lfloor b\rfloor$ belongs to the set $\{\lfloor a
\rfloor, \lfloor a \rfloor\pm 1\}$, which implies that $\tilde e_m
\in\{0, \pm 1\}$. This completes the proof.

\end{document}